\documentclass[useAMS,usenatbib]{mn2e}
\usepackage{amsmath}
\usepackage{amssymb}
\usepackage{aas_macros}
\usepackage{natbib}
\usepackage{times}
\usepackage{graphicx}
\usepackage{bm}

\newlength{\colwidth}
\newcommand{\cm}{{\rm cm}}
\newcommand{\g}                         {{\rm g}}

\newcommand{\Myr}                      {\,{\rm Myr}}

\newcommand{\pc}                        {\,{\rm pc}}

\newcommand{\pkpc}                      {\,{\rm pkpc}}

\newcommand{\ckpc}                      {\,{\rm ckpc}}
\newcommand{\cMpc}                      {\,{\rm cMpc}}
\newcommand{\Msun}                    {\,{\rm M}_\odot}

\newcommand{\hMsun}                  {\,h^{-1}\,{\rm M}_\odot}
\newcommand{\kms}                      {\,\,{\rm km}\,\,{\rm s}^{-1}}
\newcommand{\cmcubed}              {{\rm cm}^{-3}}
\newcommand{\Zsun}                     {\,{\rm Z}_\odot}
\newcommand{\erg}                       {\,{\rm erg}}

\newcommand{\K}                          {\,{\rm K}}

\newcommand{\Msolyrkpcsq}{{\rm M}_\odot\,{\rm yr}^{-1}\,{\rm kpc}^{-2}}
\newcommand{\nHtc}{n_{{\rm H},t_{\rm c}}}

\newcommand{\gadget}         {\textsc{Gadget3}}

\newcommand{\cloudy}         {\textsc{Cloudy}}

\newcommand{\nH}           {n_{\rm H}}

\newcommand{\mockalph}[1]{}

\title[EAGLE: calibration and model variations]{The EAGLE simulations of galaxy formation: calibration of subgrid physics and model variations}
 \author[R.A.Crain et al.]  {\parbox[h]{160mm}{ 
    Robert A. Crain$^{1,2}$\thanks{E-mail: r.a.crain@ljmu.ac.uk}, Joop Schaye$^1$, Richard G. Bower$^3$, Michelle Furlong$^3$, Matthieu Schaller$^3$, Tom Theuns$^{3}$, Claudio Dalla Vecchia$^{4,5}$, Carlos S. Frenk$^3$, Ian G. McCarthy$^2$, John C. Helly$^3$, Adrian Jenkins$^3$, Yetli M. Rosas-Guevara$^6$, Simon D. M. White$^7$ and James W. Trayford$^3$}
  \vspace{6pt}\\
  $^1$Leiden Observatory, Leiden University, PO Box 9513, 2300 RA Leiden, the Netherlands\\
  $^2$Astrophysics Research Institute, Liverpool John Moores University, 146 Brownlow Hill, Liverpool, L3 5RF\\
  $^3$Institute for Computational Cosmology, Department of Physics, University of Durham, South Road, Durham, DH1 3LE\\
  $^4$Instituto de Astrof\'isica de Canarias, C/V\'ia L\'actea s/n, 38205 La Laguna, Tenerife, Spain\\
  $^5$Departamento de Astrof\'isica, Universidad de La Laguna, Av. del Astrof\'isico Franciso S\'anchez s/n, 38206 La Laguna, Tenerife, Spain\\
  $^6$CRAL, Observatoire de Lyon, Universit\'e Lyon 1, 9 Avenue Ch. Andr\'e, F-69561 Saint Genis Laval Cedex, France\\
  $^7$Max-Planck-Institut f\"ur Astrophysik, Karl-Schwarzschild-Str. 1, D-85748 Garching, Germany}

\begin{document}

\date{\today}
\pagerange{\pageref{firstpage}--\pageref{lastpage}} \pubyear{2011}

\maketitle

\label{firstpage}

\begin{abstract}
We present results from thirteen cosmological simulations that explore
the parameter space of the ``Evolution and Assembly of GaLaxies and
their Environments'' (EAGLE) simulation project. Four of the
simulations follow the evolution of a periodic cube $L=50\cMpc$ on a
side, and each employs a different subgrid model of the energetic
feedback associated with star formation. The relevant parameters were
adjusted so that the simulations each reproduce the observed galaxy
stellar mass function at $z=0.1$. Three of the simulations fail
to form disc galaxies as extended as observed, and we show
analytically that this is a consequence of numerical radiative losses
that reduce the efficiency of stellar feedback in high-density gas.
Such losses are greatly reduced in the fourth simulation - the EAGLE
reference model - by injecting more energy in higher density gas. This model produces galaxies with the observed size distribution, and also reproduces many galaxy scaling relations. In
the remaining nine simulations, a single parameter or process of the reference
model was varied at a time. We find that the properties of galaxies with stellar mass $\lesssim M^\star$ (the ``knee'' of the galaxy stellar mass function) are largely governed by feedback
associated with star formation, while those of more massive galaxies
are also controlled by feedback from accretion onto their central
black holes. Both processes must be efficient in order to reproduce
the observed galaxy population. In general, simulations that
have been calibrated to reproduce the low-redshift galaxy stellar mass function will still not
form realistic galaxies, but the additional requirement that galaxy
sizes be acceptable leads to agreement with a large range of
observables.
\end{abstract}


\section{Introduction}
\label{sec:introduction}

The formation, assembly and evolution of cosmic structures is orchestrated by gravitational collapse. The non-linearity of this process precludes a fully-predictive analytic theory of structure formation, requiring that the confrontation of theoretical expectations with observational measurements must generally proceed via numerical simulations. The predictions of cosmological simulations based on the prevailing $\Lambda$-cold dark matter ($\Lambda$CDM) paradigm, in the regime where those outcomes are determined primarily by gravitational forces, have been corroborated by a diverse range of observational tests. These include, but are not limited to, cosmic shear induced by large-scale structure \citep[e.g.][]{Fu_et_al_08_short}, the abundance of brightest cluster galaxies \citep[BCGs, e.g.][]{Rozo_et_al_10_short}, tests of the cosmic expansion rate \citep[e.g.][]{Blake_et_al_11a_short} and the distance-redshift relation \citep[e.g.][]{Blake_et_al_11b_short}, redshift-space distortions of the 2-point correlation function \citep[e.g.][]{Beutler_et_al_12_short} and the luminosity-distance relation of type Ia supernovae \citep[e.g.][]{Suzuki_et_al_12_short}. 

The formation and evolution of galaxies is governed ultimately, however, by the interaction of the diverse physical processes that, in addition to gravity, influence baryonic matter. The inclusion of these processes in simulations is recognised as a major challenge, owing both to the complexity of the physical processes, and the difficulty of developing numerical algorithms able to accurately model their effects in a computationally efficient manner. This challenge has, by and large, impeded cosmological hydrodynamical simulations from yielding galaxy populations whose properties are consistent with observational measurements. Although imperfect models can prove instructive, greater confidence is generally ascribed to those that more accurately resemble the observed Universe. Moreover, the reproduction of key observables is often a prerequisite for testing particular aspects of galaxy formation theory. For example, when wishing to study the evolution of angular momentum in disc galaxies, a model that reproduces their observed size and rotation velocity is clearly desirable.

The reproduction of these particular diagnostics has in fact become a cause c\'{e}l\`{e}bre for the simulation community, owing to the long-standing need to address the closely related ``overcooling'' \citep{Cole_91,White_and_Frenk_91,Blanchard_Valls_Gabaud_and_Mamon_92,Balogh_et_al_01} and ``angular momentum'' \citep{Katz_and_Gunn_91,Navarro_and_White_94} problems. In the absence of feedback, gas efficiently radiates the heat it acquires from thermalising its gravitational potential. This excess dissipation has two principal consequences: i) the fraction of gas that is converted into stars by the present epoch is much higher than observed, and ii) the formation and coalescence of dense clumps spuriously drains angular momentum from the baryons. Simulated galaxies therefore form too many stars (and do so too early), they are more compact than observed, and they exhibit insufficient rotational support. The inclusion of prescriptions for energetic feedback processes in models has been shown to alleviate these problems \citep{Abadi_et_al_03a,Sommer_Larsen_Gotz_and_Portinari_03,Springel_and_Hernquist_03a}, and has enabled several groups to conduct simulations of the $\Lambda$CDM cosmogony that form galaxies with sizes and rotation curves that are, for particular galaxy masses, consistent with observational measurements \citep[e.g.][]{Governato_et_al_04_short,Okamoto_et_al_05,Sales_et_al_10,Guedes_et_al_11,McCarthy_et_al_12,Brook_et_al_12,Munshi_et_al_13_short,Aumer_et_al_13,Marinacci_Pakmor_and_Springel_14}. 

In spite of this success, the detailed behaviour of the multiphase interstellar medium (ISM) when subject to energetic feedback remains ill-understood, and the community has yet to converge on unique solutions to the overcooling and angular momentum problems \citep{Scannapieco_et_al_12_short}. The principal uncertainty is arguably one of accounting. Firstly, it is not known what are the energy, momentum and mass fluxes incident upon the ISM and star-forming complexes therein \citep[but see][]{Lopez_et_al_11,Rosen_et_al_14}, due to mechanisms such as radiation pressure and winds from O-class stars, asymptotic giant branch (AGB) stars and active galactic nuclei (AGN); the photoionisation and photoelectric heating of HII regions by radiation associated with stars (including X-ray binaries) and the accretion discs of black holes (BHs); and thermonuclear and core collapse supernovae (SNe). A second, often overlooked issue is that it is unknown what fraction of the incident energy is dissipated by radiative processes and thermal conduction \citep[e.g.][]{Orlando_et_al_05}, and what fraction of the incident momentum is lost due to cancellation. Estimating these initial ``losses'' is a long-standing problem in the study of the ISM, not least because of the extreme resolution and dynamic range demands of the problem: the losses are typically established on scales significantly smaller than a parsec \citep[e.g.][]{Mellema_Kurk_and_Rottgering_02,Fragile_et_al_04,Yirzak_Frank_and_Cunningham_10}. This is at least three orders of magnitude smaller than the typical size of an ISM resolution element in simulations of large cosmological volumes. 

Since these losses cannot be modelled directly by cosmological simulations, their impact on resolved scales must be incorporated into phenomenological ``subgrid'' treatments that approximate the action of unresolved processes, and couple them to resolved scales\footnote{We refer to losses on these scales as ``subgrid losses''. Losses induced by processes acting on scales that are resolved by cosmological simulations can also be significant, and dependent upon the subgrid implementation; we term these ``macroscopic losses''.}. The implementation and parametrisation of subgrid routines is therefore the greatest source of uncertainty in cosmological simulations, and adjustment of these characteristics can result in the dramatic alteration of simulation outcomes \citep{Okamoto_et_al_05,Schaye_et_al_10_short,Haas_et_al_13a,Haas_et_al_13b,Scannapieco_et_al_12_short,Vogelsberger_et_al_13,Le_Brun_et_al_14,Torrey_et_al_14}. Until small-scale losses can be accurately computed and appropriately incorporated into subgrid routines, it will remain impossible to formulate a truly predictive cosmological simulation that can, for example, yield \textit{ab initio} estimates of the stellar mass of galaxies or the mass of their central BH.

In a companion paper, \citet[][hereafter S15]{Schaye_et_al_15_short} argue that the appropriate methodology is therefore to \textit{calibrate} the parameters of subgrid routines, in order that simulations reproduce well characterised observables. The calibrated observables themselves cannot then be advanced as predictions of the model, but those not considered during the calibration can reasonably be considered as consequences of the implemented astrophysics. An obvious advantage of this approach is that, by ensuring that key properties of the galaxy population are reproduced, simulations can be used to address the widest range of problems. A related advantage is that, since any alteration to the resolution of a calculation will in general necessitate a recalibration of the model, the adopted subgrid routines need not sacrifice physical detail in order to realise numerical convergence.

Adopting this philosophy, S15 introduced the ``Evolution and Assembly of GaLaxies and their Environments'' (EAGLE) project, a suite of cosmological hydrodynamical simulations of the $\Lambda$CDM cosmogony conducted by the Virgo Consortium\footnote{See also http://eagle.strw.leidenuniv.nl and http://icc.dur.ac.uk/Eagle}. Feedback from star formation and AGN is implemented thermally, such that outflows develop as a result of pressure gradients and without the need to impose winds `by hand', for example by specifying their velocity and mass loading with respect to the star formation rate. The parameters of the subgrid routines governing feedback associated with star formation and the growth of BHs are calibrated to reproduce the observed $z=0.1$ galaxy stellar mass function (GSMF) and the relation between the mass of galaxies and their central BH, respectively, whilst also seeking to yield galaxies with sizes (i.e. effective radius) similar to those observed. S15 focussed on the EAGLE reference model (``Ref''), and a complementary model designed to meet the calibration criteria at higher resolution (``Recal'')\footnote{S15 also introduced a third model that better reproduces the observed properties of intragroup gas at intermediate resolution by adopting a higher AGN heating temperature (``AGNdT9'').}. Besides demonstrating that cosmological simulations can be calibrated to reproduce these diagnostics successfully with unprecedented accuracy, the study showed that the simulations reproduce a diverse and representative set of low-redshift observables that were not considered during the calibration process. In a separate paper, \citet{Furlong_et_al_14_short} show that the EAGLE simulations also broadly reproduce the observed GSMF as early as $z=7$, and accurately track its evolution to the present day.

This paper introduces many more simulations from the EAGLE project. The key aims of this study are to illustrate the reasons for the parametrisation adopted by the EAGLE reference model, and the sensitivity of its outcomes to the variation of the key subgrid parameters. The simulations explored here are naturally divided into two categories. The first comprises four simulations calibrated to yield the $z=0.1$ GSMF and central BH masses as a function of galaxy stellar mass. The models, one of which is the EAGLE reference model, differ in terms of the adopted subgrid efficiency of feedback associated with star formation, and the fashion by which this efficiency depends (if at all) upon the properties of the local environment. The successful reproduction of the calibration diagnostics by each of the models highlights that these observables alone do not identify a unique ``solution'', and indicates that complementary constraints are necessary to break modelling degeneracies and, potentially, motivate the inclusion of additional complexity. 

The second category comprises simulations each featuring a variation of a single subgrid parameter value with respect to the reference model. These calculations enable an examination of the role of these parameters in a fashion similar to the OWLS project \citep{Schaye_et_al_10_short}, and highlight the sensitivity of outcomes to the variation of these parameters. In common with complementary studies, these simulations indicate that the properties of the simulated galaxy population are most sensitive to the subgrid parameters governing the efficiency of energy feedback \citep{Schaye_et_al_10_short,Scannapieco_et_al_12_short,Haas_et_al_13a,Haas_et_al_13b,Vogelsberger_et_al_13}. 

This paper is structured as follows. The simulation initial conditions, and the algorithms used to evolve them, are described in \S\ref{sec:simulations}. The parametrisation of the four models that are calibrated to reproduce the $z=0.1$ GSMF is described in \S\ref{sec:calibrated_simulations}. Results from these simulations, which serve as a motivation for the development of the reference model,  are presented in \S\ref{sec:results}, where results from simulations featuring single-parameter variations of the reference model are also shown. Finally, the results are summarised and discussed in \S\ref{sec:summary_and_discussion}.

\section{Simulations and subgrid physics}
\label{sec:simulations}

This section comprises an overview of the simulation setup and subgrid physics implementation. It includes similar information to Sections 3 \& 4 of S15, so readers familiar with the simulations may wish to skip this section. A relatively comprehensive description of the subgrid implementations of star formation and feedback is retained here, because these details are a necessary foundation for later sections.

The cosmological parameters assumed by the EAGLE simulations are those recently inferred by the \citet{Planck_2013_paperI_short,Planck_2014_paperXVI_short}, the key parameters being $\Omega_{\rm m} =  0.307$, $\Omega_\Lambda = 0.693$, $\Omega_{\rm b} = 0.04825$, $h = 0.6777$ and  $\sigma_8 =  0.8288$. Initial conditions adopting these parameters were generated using a transfer function created with the \textsc{camb} software \citep{Lewis_Challinor_and_Lasenby_00}, the $2^{\rm nd}$-order Lagrangian perturbation theory method of \citet{Jenkins_10}, and the Gaussian white noise field \textit{Panphasia} \citep{Jenkins_13,Jenkins_and_Booth_13}. A complete description of the generation of the initial conditions is provided in Appendix B of S15, and the tools necessary to generate them independently are available online\footnote{See http://eagle.strw.leidenuniv.nl.}. 

The simulations were evolved by a modified version of the $N$-body TreePM smoothed particle hydrodynamics (SPH) code \gadget, last described by \citet{Springel_05}. The modifications comprise
updates to the hydrodynamics algorithm and the time-stepping criteria, and the addition of subgrid routines governing the phenomenological implementation of processes occurring on scales below the resolution limit of the simulations. The updates to the hydrodynamics algorithm, which we collectively refer to as ``Anarchy'' (Dalla Vecchia in prep.), comprise an implementation of the pressure-entropy formulation of SPH derived by \citet{Hopkins_13}, the artificial viscosity switch proposed by \citet{Cullen_and_Dehnen_10}, an artificial conduction switch similar to that proposed by \citet{Price_08}, the $C^2$ smoothing kernel of \citet{Wendland_95}, and the time-step limiter of \citet{Durier_and_Dalla_Vecchia_12}. 

The subgrid routines represent an evolution of those used for the GIMIC \citep{Crain_et_al_09_short}, OWLS \citep{Schaye_et_al_10_short} and cosmo-OWLS \citep{Le_Brun_et_al_14} projects, and include element-by-element radiative cooling and photoionisation heating for 11 species, star formation, stellar mass loss, energy feedback from star formation, gas accretion onto and mergers of BHs, and AGN feedback.
The key updates with respect to the routines used by OWLS are the inclusion of a metallicity dependence in the star formation law, the implementation of energy feedback associated with star formation via stochastic thermal heating, and the inclusion of a viscous transport limit on the BH accretion rate. 

S15 introduced the resolution nomenclature of the EAGLE project. ``Intermediate-resolution'' simulations have particle masses corresponding to an $L=100~{\rm comoving~Mpc}$ (hereafter~$\cMpc$) volume realised with $2 \times 1504^3$ particles (an equal number of baryonic and dark matter particles), such that the initial gas particle mass is $m_{\rm g}=1.81 \times 10^6\Msun$, and the mass of dark matter particles is $m_{\rm dm}=9.70 \times 10^6\Msun$. The Plummer-equivalent gravitational softening length is fixed in comoving units to $1/25$ of the mean interparticle separation ($2.66~{\rm comoving~kpc}$, hereafter~$\ckpc$) until $z=2.8$, and in proper units ($0.70~{\rm proper~kpc}$, hereafter~$\pkpc$) at later times. The intermediate-resolution simulations marginally resolve the Jeans scales at the star formation threshold ($n_{\rm H}\simeq 10^{-1}\,\cmcubed$) in the warm ($T\simeq 10^4\K$) ISM. ``High-resolution'' simulations adopt particle masses and softening lengths that are smaller by factors of eight and two, respectively. The SPH kernel size, specifically its support radius, is limited to a minimum of one-tenth of the  gravitational softening scale. This study focusses on intermediate-resolution simulations using volumes of side $L=25$, $50$ and $100\cMpc$, which therefore comprise $2 \times 376^3$, $2 \times 752^3$ and $2 \times 1504^3$ particles, respectively. 

Galaxies and their host haloes are identified by a multi-stage process, beginning with the application of the friends-of-friends (FoF) algorithm \citep{DEFW85} to the dark matter particle distribution, with a linking length of $b=0.2$ times the mean interparticle separation. Gas, star and BH particles are associated with the FoF group, if any, of their nearest neighbour dark matter particle. The SUBFIND algorithm \citep{Springel_et_al_01,Dolag_et_al_09} is then used to identify self-bound substructures, or \textit{subhaloes}, within the full particle distribution (gas, stars, BHs and dark matter) of FoF haloes. The subhalo comprising the particle with the minimum gravitational potential, which is almost exclusively the most massive subhalo, is defined as the central subhalo, the remainder being satellite subhaloes. The coordinate of the particle with the minimum potential also defines the position of the halo, about which is computed the spherical overdensity \citep[SO;][]{Lacey_and_Cole_94} mass, $M_{200}$, for the adopted density contrast of $200$ times the critical density, $\rho_{\rm c}$. Satellite subhaloes separated from their central galaxy by less than the minimum of $3\pkpc$ and the stellar half-mass radius of the central galaxy are merged into the latter; this step eliminates a small number of low-mass subhaloes dominated by single, high-density gas particles or BHs. Finally, when quoting the properties of galaxies (e.g. stellar mass, star formation rate), only those subhalo particles within a spherical aperture of radius $30\pkpc$ are considered. S15 (their Figure 6) demonstrated that this practice yields stellar masses comparable to those recovered within a projected circular aperture with the Petrosian radius at $z=0.1$.

\subsection{Radiative processes}
Radiative cooling and heating rates are computed on an element-by-element basis by interpolating tables, generated with \cloudy\ \citep[version 07.02,][]{Ferland_et_al_98}, that specify cooling rates as a function of density, temperature and redshift, under the assumption that the gas is optically thin, is in ionisation equilibrium, and is exposed to the cosmic microwave background and a spatially-uniform, temporally-evolving \citet{Haardt_and_Madau_01} UV/X-ray background \citep[for further details, see][]{Wiersma_Schaye_and_Smith_09}. The UV/X-ray background is imposed instantaneously at $z=11.5$. To account for enhanced photoheating rates (relative to the optically thin rates assumed here) during the epochs of reionisation, $2\,{\rm eV}$ per proton mass is injected, rapidly heating gas to $\sim10^4\K$. This is done instantaneously at $z=11.5$ (consistent with \textit{Planck} constraints) for HI reionisation, but for HeII the energy injection is distributed in redshift with a Gaussian function centred about $z=3.5$ with a width of $\sigma(z)=0.5$. This ensures that the thermal evolution of the intergalactic medium mimics that inferred by \citet{Schaye_et_al_00}.

\subsection{The ISM and star formation} 
\label{sec:ism_and_sf}

Simulations of large cosmological volumes lack, in general, the resolution and physics to model the cold ($T \ll 10^4\K$) interstellar gas phase from which molecular clouds and stars form. A global temperature floor, $T_{\rm eos}(\rho)$ is therefore imposed, corresponding to a polytropic equation of state, $P_{\rm eos} \propto \rho^{\gamma_{\rm eos}}$, normalised to $T_{\rm eos} = 8000\K$ at $n_{\rm H} = 0.1\,\cmcubed$. A fiducial polytrope of $\gamma_{\rm eos}=4/3$ is adopted, since this ensures that the Jeans mass, and the ratio of the Jeans length to the SPH support radius, are independent of density, thus inhibiting spurious fragmentation \citep{Schaye_and_Dalla_Vecchia_08}. The effect of varying $\gamma_{\rm eos}$ is explored in \S\,\ref{sec:reference_variations}, where simulations conducted using isothermal ($\gamma_{\rm eos}=1$) and adiabatic ($\gamma_{\rm eos}=5/3$) equations of state are examined.

A second temperature floor of $8000\K$ is imposed for gas with $\nH > 10^{-5}\,\cmcubed$, which prevents metal-rich gas from cooling to very low temperatures, since the physical processes required to model dense, low-temperature gas are not included here. This floor does not apply to low-density (i.e. intergalactic) gas, since such gas cools adiabatically and is modelled accurately by the hydrodynamics scheme.

Star formation is implemented stochastically, based on the pressure law scheme of \citet{Schaye_and_Dalla_Vecchia_08}. Under the (reasonable) assumption that star-forming gas is self-gravitating, the observed Kennicutt-Schmidt star formation law \citep{Kennicutt_review_98},
\begin{equation}
\dot{\Sigma}_\star = A \left( \frac{\Sigma_{\rm g}}{1~\Msun \pc^{-2}}\right)^n,
\end{equation}
where $\Sigma_\star$ and $\Sigma_{\rm g}$ are the surface density density of stars and gas, respectively, can be expressed as:
\begin{equation}
\dot{m}_\ast = m_{\rm g} A \left (1~\Msun \pc^{-2}\right)^{-n} \left ({\frac{\gamma}{G}} f_{\rm g} P\right )^{(n-1)/2},
\label{eq:sflaw}
\end{equation}
where $m_{\rm g}$ is the gas particle mass, $\gamma=5/3$ is the ratio of specific heats (and should not be confused with $\gamma_{\rm eos}$), $G$ is the gravitational constant, $f_{\rm g}$ is the mass fraction in gas (assumed to be unity), and $P$ is the total pressure. This pressure law implementation is advantageous for two reasons. Firstly, the free parameters of the star formation law ($A$, $n$) are specified explicitly by observations: the values $A=1.515\times10^{-4}~\Msolyrkpcsq$ and $n=1.4$ ($n=2$ for $n_{\rm H} > 10^3\,\cmcubed$) are adopted, where the value of $A$ has been adjusted from that reported by \citet{Kennicutt_review_98} to convert from the Salpeter initial stellar mass function (IMF) to the \citet{Chabrier_03} form adopted by the simulations. Secondly, this implementation guarantees that the observed Kennicutt-Schmidt relation is reproduced for any equation of state (i.e. any combination of $T_{\rm eos}$ and $\gamma_{\rm eos}$) applied to star-forming gas. This is in contrast to volumetric star formation laws, which must be recalibrated whenever the equation of state is altered. 

Star formation occurs only in cold, dense gas, requiring that a density threshold for star formation, $n_{\rm H}^\star$, be imposed. Since the transition from a warm, neutral phase to a cold, molecular one occurs at lower densities and pressures in metal-rich (and hence dust-rich) gas, we adopt the metallicity-dependent star formation threshold proposed by \citet{Schaye_04}, which was implemented in the OWLS simulation ``SFTHRESHZ'':
\begin{equation} 
n_{\rm H}^\star(Z)=\min \left [ 0.1\,\cmcubed \left ({\frac{Z}{0.002}}\right )^{-0.64}, 10\,\cmcubed \right]
\label{eq:sfthreshz} 
\end{equation} 
where $Z$ is the gas metallicity. Hydrogen number density, $n_{\rm H}$, is related to the overall gas density, $\rho$, via $n_{\rm H} \equiv X \rho / m_{\rm H}$, where $X$ is the hydrogen mass fraction and $m_{\rm H}$ is the mass of a hydrogen atom. To examine the effects, if any, of adopting this metallicity-dependent threshold, the EAGLE suite includes a simulation that adopts a constant threshold of $n_{\rm H}^\star = 0.1\,\cmcubed$, which was the fiducial approach of OWLS. In both cases, to prevent star formation in low-overdensity gas at high redshift, star-forming gas is required to have an overdensity $\delta > 57.7$.

\subsection{Stellar evolution and mass loss}  

The implementation of stellar evolution and mass loss is based upon that described by \citet{Wiersma_et_al_09}. Star particles are treated as simple stellar populations (SSPs) with an IMF of the form proposed by \citet{Chabrier_03}, spanning the range $0.1-100~\Msun$. At each time step and for each stellar particle, those stellar masses reaching the end of the main sequence phase are identified using metallicity-dependent lifetimes, and the fraction of the initial particle mass reaching this evolutionary stage is used, together with the particle initial elemental abundances and nucleosynthetic yield tables, to compute the mass of each element that is lost through winds from AGB stars, winds from massive stars, and type II SNe. Eleven elements are tracked, and the mass and energy lost through type Ia SNe is also computed, assuming the rate of type Ia SNe per unit stellar mass is specified by an empirically-motivated exponential delay function. 

\subsection{Energy feedback from star formation}
\label{sec:sf_feedback}

Stars inject energy and momentum into the ISM via stellar winds, radiation, and SNe. These processes are particularly important for massive, short-lived stars and, if star formation is sufficiently vigorous, the associated feedback can drive large-scale galactic outflows \citep[e.g.][]{Veilleux_Cecil_and_Bland-Hawthorn_05}. At present, simulations of large cosmological volumes lack the resolution necessary to model the self-consistent development of outflows from feedback injected on the scales of individual star clusters, and must appeal to a subgrid treatment.

In the simplest implementation of energy feedback by thermal heating, the energy produced at each timestep by a star particle is distributed to a number of its neighbouring hydrodynamic resolution elements, supplementing their internal energy. \citet[][see also \citealt{Dalla_Vecchia_and_Schaye_08}, \citealt{Creasey_et_al_11}, \citealt{Keller_et_al_14} and \citealt{Creasey_Theuns_and_Bower_15}]{Dalla_Vecchia_and_Schaye_12} argue that the feedback energy (canonically $\sim10^{51}\,\erg$ per $100\,{\rm M}_\odot$ for a standard IMF if considering SNe as the sole energy source) is initially distributed over too much mass: the mass of at least one resolution element, $\mathcal{O}(10^4-10^7)\Msun$, rather than that of the actual ejecta,  $\mathcal{O}(10^0-10^1)\Msun$. The resulting temperature increment is then far smaller than in reality, and by extension the radiative cooling time of the heated gas is much too short. Pressure gradients established by the heating are too shallow and, perhaps more importantly, are typically erased on a (radiative) timescale shorter than the sound crossing time of a resolution element.

The EAGLE simulations adopt the stochastic thermal feedback scheme of \citet{Dalla_Vecchia_and_Schaye_12}, in which the temperature increment, $\Delta T_{\rm SF}$, of heated resolution elements is specified. Besides enabling one to mitigate the problem described above, a stochastic implementation of feedback is advantageous because it enables the quantity of energy injected \textit{per feedback event} to be specified, even if the mean quantity of energy injected per unit stellar mass formed is fixed. Having specified $\Delta T_{\rm SF}$, the probability that a resolution element neighbouring a young star particle is heated, is determined by the fraction of the energy budget that is available for feedback\footnote{\citet{Durier_and_Dalla_Vecchia_12} show that implementations of kinetic and thermal feedback converge to similar solutions in the limit that the cooling time is long. We adopt a thermal implementation here for consistency with our AGN feedback implementation, and because numerical tests indicate that it is less susceptible to problems stemming from poor numerical sampling of outflows.}. For consistency with the nomenclature introduced by \citet{Dalla_Vecchia_and_Schaye_12}, we refer to this fraction as $f_{\rm th}$.

We adopt the convention that $f_{\rm th}=1$ equates to an expectation value of the injected energy of $1.736\times 10^{49}\,\erg\,\Msun^{-1}$ ($8.73\times 10^{15}\,\erg\,\g^{-1}$) of stellar mass formed. This corresponds to the energy available from type II SNe resulting from a Chabrier IMF, subject to two assumptions. Firstly, that $6-100\Msun$ stars are the progenitors of type II SNe ($6-8\Msun$ stars explode as electron capture supernovae in models with convective overshoot; e.g.\ \citealt{Chiosi_Bertelli_and_Bressan_92}), and secondly that each SN liberates $10^{51}\erg$. We inject energy once for each star particle, when it reaches an age of $30\Myr$.

For thermal feedback to be effective, the pressure gradient established by the heating must be able to perform work on the gas (via sound waves, or shocks for supersonic flows) on a timescale that is shorter than that required to erase the gradient via radiative cooling. By comparing the sound crossing and cooling timescales for heated resolution elements, \citet{Dalla_Vecchia_and_Schaye_12} derived an estimate for the maximum gas density, $\nHtc$, at which their stochastic heating scheme can be efficient (their equation 18),
\begin{equation}
\nHtc \sim 10~\cm^{-3} \left (\frac{T}{10^{7.5}\,\K}\right )^{3/2} \left (\frac{m_{\rm g}}{10^6\,\Msun}\right)^{-1/2}, 
\label{eq:nhtc}
\end{equation}
where $T > \Delta T_{\rm SF}$ is the post-heating temperature. For simplicity, the cooling is assumed to be Bremsstrahlung-dominated, so the value of $\nHtc$ will be over-estimated in the temperature regime where (metal-)line cooling dominates ($T\ll10^7\K$). S15 noted that some stars do form in the EAGLE simulations from gas with density greater than the critical value, in which case the spurious (numerical) radiative losses are significant. In the case of such losses being significant, the energy budget used to reproduce the GSMF in the simulation will likely be an overestimate of that required in Nature. 

Equation \ref{eq:nhtc} indicates that numerical losses associated with stars forming from high-density gas can be mitigated by appealing to a higher heating temperature, $\Delta T_{\rm SF}$. However, this is not an ideal solution. For a fixed quantity of feedback energy per stellar mass formed (i.e. constant $f_{\rm th}$), the probability that a star particle triggers a heating event is inversely proportional to $\Delta T_{\rm SF}$. Based on the energy budget described above, the expectation value of the number of resolution elements (in this case, SPH particles) heated by a star particle is specified by \citet[][equation 8]{Dalla_Vecchia_and_Schaye_12} as:
\begin{equation}
\left < N_{\rm heat}\right > \approx 1.3 f_{\rm th} \left (\frac{\Delta T_{\rm SF}}{10^{7.5}\,\K}\right )^{-1}.
\label{eq:Nav}
\end{equation}
In the regime of $\Delta T_{\rm SF} \gg 10^{7.5}\K$, the probability of heating a single resolution element becomes small, leading to poor sampling of the feedback cycle. We therefore only consider models adopting $\Delta T_{\rm SF} = 10^{7.5}\K$. 

If $\Delta T_{\rm SF}$ is sufficiently high to ensure that numerical losses are small, the \textit{physical} efficiency of feedback can be controlled by adjusting $f_{\rm th}$. This mechanism is a means of modelling the subgrid radiative losses that are not addressed by our simple treatment of the ISM. Because these losses should depend on the physical conditions in the ISM, there is physical motivation to specify $f_{\rm th}$ as a function of the local properties of the gas. Primarily, it is the freedom to adjust $f_{\rm th}$ that enables the simulations to be calibrated.

\subsection{Black holes and AGN feedback} 
\label{sec:bh}

Feedback associated with the growth of BHs is an essential ingredient of the EAGLE simulations. Besides regulating the growth of the BHs, AGN feedback quenches star formation in massive galaxies and shapes the gas profiles of their host haloes. The implementation adopted here consists of two elements, namely i) a prescription for seeding galaxies with BHs and for following their growth via mergers and gas accretion, and ii) a prescription for coupling the radiated energy, liberated by BH growth, to the ISM. The method for the former is based on that introduced by \citet{Springel_Di_Matteo_and_Hernquist_05} and modified by \citet{Booth_and_Schaye_09} and \citet{Rosas_Guevara_et_al_14_short}, while the method for the latter is similar to that described by \citet{Booth_and_Schaye_09}. 

Following \citet{Springel_Di_Matteo_and_Hernquist_05}, seed BHs are placed at the centre of every halo more massive than $10^{10}\,\Msun/h$ that does not already contain a BH. Candidate haloes are identified by running a FoF algorithm with linking length $b=0.2$ on the dark matter distribution. When a seed is required, the highest density gas particle is converted into a collisionless BH particle, inheriting the particle mass and acquiring a subgrid BH mass $m_{\rm BH} = 10^5\,\Msun/h$ which, for the intermediate resolution simulations considered here, is smaller than the initial gas particle mass by a factor of $12.3$. Calculations of BH properties are therefore functions of $m_{\rm BH}$, whilst gravitational interactions are computed using the particle mass. When the subgrid BH mass exceeds the particle mass, BH particles stochastically accrete neighbouring gas particles such that particle and subgrid BH masses grow in concert. BHs are prevented from ``wandering'' out of their parent haloes by forcing those with mass $< 100 m_{\rm g}$ to migrate towards the position of the minimum of the gravitational potential in their halo. 

BHs are merged if separated by a distance that is smaller than both the BH kernel size, $h_{\rm BH}$, and three gravitational softening lengths, and if their relative velocity is smaller than the circular velocity at the distance $h_{\rm BH}$, $v_{\rm rel} < \sqrt{G m_{\rm BH}/h_{\rm BH}}$, where $h_{\rm BH}$ and $m_{\rm BH}$ are, respectively, the SPH kernel size and the subgrid mass of the most massive BH in the pair. The relative velocity threshold prevents BHs from merging during the initial stages of galaxy mergers.

\subsubsection{Gas accretion onto black holes}
\label{sec:bh_accretion}

The gas accretion rate, $\dot{m}_{\rm accr}$, is specified by the minimum of the Eddington rate,
\begin{equation}
\dot{m}_{\rm Edd} = \frac{4\pi G m_{\rm BH} m_{\rm p}}{\epsilon_{\rm r} \sigma_{\rm T} c},
\label{eq:mdotEdd}
\end{equation}
and
\begin{equation}
\dot{m}_{\rm accr} = \min\left (\dot{m}_{\rm Bondi} [(c_{\rm s}/V_\phi)^3/C_{\rm visc}], 
\dot{m}_{\rm Bondi}\right ), 
\label{eq:mdotaccr}
\end{equation}
where $\dot{m}_{\rm Bondi}$ is the Bondi-Hoyle (\citeyear{Bondi_and_Hoyle_44}) rate applicable to spherically symmetric accretion, 
\begin{equation}
\dot{m}_{\rm Bondi} = \frac{4\pi G^2 m_{\rm BH}^2 \rho}{(c_{\rm s}^2 + v^2)^{3/2}}.
\label{eq:bondihoyle}
\end{equation}
Here $m_{\rm p}$ is the proton mass, $\sigma_{\rm T}$ the Thomson cross section, $c$ the speed of light, $\epsilon_{\rm r}$ the radiative efficiency of the accretion disc, and $v$ the relative velocity of the BH and the gas. Finally, $V_\phi$ is the circulation speed of the gas around the BH computed using equation 16 of \citet{Rosas_Guevara_et_al_14_short} and $C_{\rm visc}$ is a free parameter related to the viscosity of a notional subgrid accretion disc. The growth of the BH is specified by 
\begin{equation} 
\dot{m}_{\rm BH} = (1-\epsilon_{\rm r}) \dot{m}_{\rm accr}. 
\end{equation}
We assume a radiative efficiency of $\epsilon_{\rm r}=0.1$. The factor $(c_{\rm s}/V_\phi)^3/C_{\rm visc}$ by which the Bondi rate is multiplied in equation \ref{eq:mdotaccr} is equivalent to the ratio of the Bondi and viscous time scales (see \citealt[][]{Rosas_Guevara_et_al_14_short}). The critical ratio of $V_\phi/c_{\rm s}$ above which angular momentum is assumed to suppress the accretion rate scales as $C_{\rm visc}^{-1/3}$. Larger values of $C_{\rm visc}$ therefore correspond to a lower subgrid kinetic viscosity, and so act to \textit{delay} the growth of BHs by gas accretion and, by extension, the onset of quenching by AGN feedback. 

\subsubsection{AGN feedback}
\label{sec:agn_feedback}

A single mode of AGN feedback is adopted, whereby energy is injected thermally and stochastically, in a manner analogous to energy feedback from star formation (see \S\,\ref{sec:sf_feedback}). The energy injection rate is $\epsilon_{\rm f} \epsilon_{\rm r} \dot{m}_{\rm accr} c^2$, where $\epsilon_{\rm f}$ is the fraction of the radiated energy that couples to the ISM. In common with the efficiency of feedback associated with star formation, $f_{\rm th}$, the value of $\epsilon_{\rm f}$ must be chosen by calibrating to observations. Because of self-regulation, the value of $\epsilon_{\rm f}$ \emph{only} affects the masses of BHs \citep{Booth_and_Schaye_09}, which vary inversely with $\epsilon_{\rm f}$, and it has little effect on the stellar mass of galaxies (provided its value is non-zero). The parameter $\epsilon_{\rm f}$ can be calibrated by ensuring the normalisation of the observed relation between BH mass and stellar mass is reproduced at $z=0$. Although implemented as a single heating mode, \citet{McCarthy_et_al_11} demonstrate that this scheme mimics quiescent `radio'-like and vigorous `quasar'-like AGN modes when the BH accretion rate is a small ($\ll 1$) or large ($\sim 1$) fraction of the Eddington rate, respectively.

S15 demonstrated that the efficiency adopted by OWLS, $\epsilon_{\rm f}=0.15$, remains a suitable choice at the higher resolution of EAGLE. Therefore, a fraction $\epsilon_{\rm f}\epsilon_{\rm r}=0.015$ of the accreted rest mass energy is coupled to the local ISM as feedback. Each BH maintains a ``reservoir'' of feedback energy, $E_{\rm BH}$. After each time step $\Delta t$, an energy $\epsilon_{\rm f} \epsilon_{\rm r} \dot{m}_{\rm accr} c^2 \Delta t$ is added to the reservoir. Once a BH has stored sufficient energy to heat at least one fluid element of mass $m_{\rm g}$, it becomes eligible to heat, stochastically, its SPH neighbours by increasing their temperature by $\Delta T_{\rm AGN}$. For each neighbour the heating probability is
\begin{equation}  
P = \frac{E_{\rm BH} }{ \Delta\epsilon_{\rm AGN} N_{\rm ngb} \left <m_{\rm g}\right > },
\end{equation} 
where $\Delta\epsilon_{\rm AGN}$ is the change in internal energy per unit mass corresponding to the temperature increment, $\Delta T_{\rm AGN}$ (the parameter $\Delta T_{\rm AGN}$ is converted into $\Delta\epsilon_{\rm AGN}$ assuming a fully ionised gas of primordial composition), $N_{\rm ngb}$ is the number of gas neighbours of the BH and $\left <m_{\rm g}\right >$ is their mean mass. The value of $E_{\rm BH}$ is then reduced by the expectation value of the injected energy. The time step of the BHs is limited to aim for probabilities $P<0.3$.

Larger values of $\Delta T_{\rm AGN}$ yield more energetic feedback events, generally resulting in reduced radiative losses (as per equation \ref{eq:nhtc}). However, larger values also make the feedback more intermittent. In general, the ambient density of gas local to the central BH of galaxies is greater than that of star-forming gas distributed throughout their discs, so a higher heating temperature is required to minimise numerical losses\footnote{BHs are in principle able to inject feedback energy at all times, unlike star particles which inject prompt feedback only once. The AGN feedback cycle can therefore remain well sampled for higher heating temperatures than is the case for star formation feedback, as long as the interval between heating events is shorter than a Salpeter time \citep{Booth_and_Schaye_09}.}. The EAGLE reference model presented by S15 adopts $\Delta T_{\rm AGN} = 10^{8.5}\,\K$. In that study an alternative intermediate-resolution model was also presented (model ``AGNdT9'') which, by appealing to $\Delta T_{\rm AGN} = 10^{9}\,\K$, was found to more accurately reproduce the observed gas fractions and X-ray luminosities of galaxy groups. This higher temperature increment was also found to be necessary in high-resolution simulations, since they resolve higher ambient densities close to BHs and hence exhibit higher cooling rates.

The values of the relevant subgrid parameters adopted by all simulations featured in this study are listed in Table \ref{tbl:simulations}.

\section{Calibrated simulations}
\label{sec:calibrated_simulations}

\begin{table*} 
\begin{center}
\begin{tabular}{l l l l l l l l l l l l}
\hline
\hline
Identifier & Side length & $N$ & $\gamma_{\rm eos}$ & $n_{\rm H}^\star$ & $f_{\rm th}$-scaling & $f_{\rm th,max}$ & $f_{\rm th,min}$ & $n_{\rm H,0}$ & $n_{\rm n}$ & $C_{\rm visc}/2\pi$ & $\Delta T_{\rm AGN}$  \\
               & [\cMpc]      &          &                               & [$\cmcubed$]    &                               &                        &                        & [$\cmcubed$] &            &                    & $\log_{10}$ [K]       \\
\hline
\textit{Calibrated models} \\
FBconst\        & 50  & 752      & $4/3$    & Eq.~\ref{eq:sfthreshz} & $-$             & $1.0$      & $1.0$       & $-$    & $-$          & $10^3$ & $8.5$ \\
FB$\sigma$      & 50   & 752      & $4/3$    & Eq.~\ref{eq:sfthreshz} & $\sigma_{\rm DM}^2$ & $3.0$      & $0.3$       & $-$    & $-$          & $10^2$ & $8.5$ \\
FBZ             & 50  & 752       & $4/3$    & Eq.~\ref{eq:sfthreshz} & $Z$             & $3.0$      & $0.3$       & $-$    & $-$          & $10^2$ & $8.5$ \\
Ref (FBZ$\rho$) & 50  & 752       & $4/3$    & Eq.~\ref{eq:sfthreshz} & $Z,\rho$        & $3.0$      & $0.3$       & $0.67$ & $2/\ln{10}$ & $10^0$ & $8.5$ \\
\hline
\textit{Ref variations} \\
eos1            & 25   & 376      & $\bm{1}$   & Eq.~\ref{eq:sfthreshz}    & $Z,\rho$  & $3.0$      & $0.3$            & $0.67$ & $2/\ln{10}$ & $10^0$ & $8.5$  \\
eos5/3          & 25  & 376       & $\bm{5/3}$ & Eq.~\ref{eq:sfthreshz}    & $Z,\rho$  & $3.0$      & $0.3$            & $0.67$ & $2/\ln{10}$ & $10^0$ & $8.5$  \\
FixedSfThresh     & 25  & 376       & $4/3$      & $\bm{0.1}$ & $Z,\rho$  & $3.0$      & $0.3$            & $0.67$ & $2/\ln{10}$ & $10^0$ & $8.5$   \\
WeakFB          & 25  & 376       & $4/3$      & Eq.~\ref{eq:sfthreshz}    & $Z,\rho$  & $\bm{1.5}$ & $\bm{0.15}$ & $0.67$ & $2/\ln{10}$ & $10^0$ & $8.5$   \\
StrongFB        & 25  & 376       & $4/3$      & Eq.~\ref{eq:sfthreshz}    & $Z,\rho$  & $\bm{6.0}$ & $\bm{0.6}$  & $0.67$ & $2/\ln{10}$ & $10^0$ & $8.5$   \\
ViscLo          & 50  & 752       & $4/3$      & Eq.~\ref{eq:sfthreshz}    & $Z,\rho$  & $3.0$      & $0.3$            & $0.67$ & $2/\ln{10}$ & $\bm{10^2}$ & $8.5$    \\
ViscHi          & 50  & 752       & $4/3$      & Eq.~\ref{eq:sfthreshz}    & $Z,\rho$  & $3.0$      & $0.3$            & $0.67$ & $2/\ln{10}$ & $\bm{10^{-2}}$ & $8.5$   \\
AGNdT8          & 50  & 752       & $4/3$      & Eq.~\ref{eq:sfthreshz}    & $Z,\rho$  & $3.0$      & $0.3$            & $0.67$ & $2/\ln{10}$ & $10^0$ & $\bm{8.0}$      \\
AGNdT9          & 50   & 752      & $4/3$      & Eq.~\ref{eq:sfthreshz}    & $Z,\rho$  & $3.0$      & $0.3$            & $0.67$ & $2/\ln{10}$ & $10^0$ & $\bm{9.0}$      \\
\hline
\end{tabular}
\caption{Parameters that are varied in the simulations. Columns are: the side length of the volume ($L$) and the particle number per species (i.e. gas, DM) per dimension ($N$), the power law slope of the polytropic equation of state ($\gamma_{\rm eos}$), the star formation density threshold ($n_{\rm H}^\star$), the scaling variable of the efficiency of star formation feedback ($f_{\rm th}$), the asymptotic maximum and minimum values of $f_{\rm th}$, the Ref model's density term denominator ($n_{\rm H,0}$) and exponent ($n_{\rm n}$) from equation \ref{eq:fth(Z,n)}, the subgrid accretion disc viscosity parameter ($C_{\rm visc}$) from equation \ref{eq:mdotaccr}, and the temperature increment of stochastic AGN heating ($\Delta T_{\rm AGN}$). The upper section comprises the four models that have been calibrated to reproduce the $z = 0.1$ GSMF, and the lower section comprises models featuring a single-parameter variations of Ref (varied parameter highlighted in bold). All models also adopt $n_{\rm Z} = 2/\ln{10}$ with the exceptions of FB$\sigma$, for which the parameter $n_{\rm Z}$ is replaced by $n_{\rm T}$ with the same numerical value (see equation \ref{eq:fth(T)}), and FBconst, for which the parameter is inapplicable.}
\label{tbl:simulations}
\end{center}
\end{table*}

As discussed by S15 (see their \S\,2), if subgrid models for energy feedback offer an incomplete description of the processes they are designed to model, are subject to numerical losses, or if the outcomes of the prescriptions are sensitive to resolution, then the true efficiencies of feedback processes cannot be predicted from first principles. It was therefore argued that cosmological hydrodynamical simulations are presently unable to yield \textit{ab initio} predictions of the stellar mass of galaxies, nor the mass of their central BH. Subgrid parameters should therefore be calibrated such that simulations reproduce desired diagnostic quantities, stellar and BH masses being germane examples.

The optimal approach to calibrating subgrid models is not unambiguous, since there can be multiple measurable outcomes that are sensitive to the adjustment of subgrid parameters, some or all of which might reasonably be considered valid constraints. For example, in the case of feedback efficiencies, one might calibrate the model to reproduce the velocity and/or mass loading of outflowing gas. However, these quantities remain ill-characterised observationally, and are sensitive to the physical scale on which they are measured (which is generally not even well known). Reproducing the properties of outflows on a particular spatial scale offers no guarantee that they are reproduced on other scales, since, for example, the interaction of outflows with the circumgalactic medium may be inadequately modelled. The choice of calibration diagnostic(s) is therefore somewhat arbitrary, but some choices can be more readily motivated. Clearly, it is necessary that any diagnostic be well characterised observationally on the scales resolved by the simulation. Perhaps the most elegant example is the star formation law which, on the $\sim 10^2\pc$ scales we follow in the ISM, can be accurately represented by the Kennicutt-Schmidt relation; as described in \S\,\ref{sec:ism_and_sf}, the free parameters of the subgrid star formation law are unambiguously prescribed by observations. In addition, it is desirable to confront calibrated models with complementary observational constraints, to minimise the risk of overlooking modelling degeneracies.

We chose to calibrate the feedback simulations to the $z=0.1$ GSMF, a practice commonly adopted by the semi-analytic galaxy formation modelling community. Low-redshift galaxy surveys enable the GSMF to be characterised in the local Universe across five decades in mass scale, with a precision that is, assuming a universal IMF, limited primarily by systematic uncertainties in the stellar evolution models used to infer the masses  \citep[][but see  \citealt{Taylor_et_al_11_short}]{Conroy_Gunn_and_White_09,Pforr_Maraston_and_Tonini_12}, peculiar velocity corrections for faint galaxies \citep[e.g.][]{Baldry_et_al_12_short}, and the method used to subtract the sky background from bright galaxies \citep[e.g.][]{Bernardi_et_al_13,Kravtsov_Vikhlinin_and_Meshscheryakov_14}. An additional motivation for appealing to the GSMF as the calibration diagnostic is that its reproduction by the simulations is a prerequisite for examination of many observable scaling relations. 

Whilst calibrating the simulations, attention was also paid to the sizes of galaxies. As with the calibration diagnostic, this choice is somewhat arbitrary, but it is readily motivated since the formation of unrealistically compact or extended galaxies would limit the utility of the simulations and likely indicate physical and/or numerical inaccuracies. The formation of disc galaxies with realistic sizes in cosmological simulations has proven to be a non-trivial challenge, leading to the identification and rectification of many shortcomings of numerical techniques \citep[e.g.][]{Sommer-Larsen_Gelato_and_Vedel_99,Ritchie_and_Thomas_01,Marri_and_White_03,Okamoto_et_al_03,Agertz_et_al_07_short,Springel_10,Hopkins_13,Hopkins_14}. In terms of the physics of galaxy assembly, \citet{Navarro_and_White_94} identified the transfer of angular momentum from cold, dense clumps of baryons to the non-dissipative dark matter residing at the outskirts of galaxy haloes as the major concern. In a comprehensive simulation comparison project, \citet{Scannapieco_et_al_12_short} highlighted that the angular momentum problem (and also the overcooling problem) remains without a consistent solution, and therefore that galaxy sizes cannot yet be uniquely predicted, even when the assembly history of their parent halo is fully specified. It cannot be assumed that the sizes of galaxy discs will be accurately reproduced by simulations, even if they successfully reproduce a suitable calibration diagnostic. For this reason, we require a model to reproduce both the GSMF and the observed size-mass relation of disc galaxies\footnote{The size evolution of both late- and early-type galaxies will be explored in detail in a forthcoming paper (Furlong et al. \textit{in prep}).} at low-redshift, in order for the calibration process to be deemed successful.

The subgrid model governing feedback associated with star formation in EAGLE is primarily dependent upon the IMF, $f_{\rm th}$ and $\Delta T_{\rm SF}$. A universal IMF is adopted throughout, and the adoption of a fixed $\Delta T_{\rm SF}=10^{7.5}\K$ was motivated in \S\,\ref{sec:sf_feedback}. Therefore, energy feedback associated with star formation is calibrated exclusively by varying $f_{\rm th}$, the fraction of the total available energy from type II SNe that couples to the ISM. The main effect on the GSMF of increasing (decreasing) $f_{\rm th}$ is to lower (raise) its normalisation in terms of the comoving number density of galaxies with stellar mass below the ``knee'' of the \citet{Schechter_76} function (demonstrated in \S\,\ref{sec:vary_fth}).

The subgrid model governing AGN feedback in EAGLE is primarily dependent upon  $\epsilon_{\rm f}$, $C_{\rm visc}$ and $\Delta T_{\rm AGN}$. The retention of the AGN feedback efficiency adopted by OWLS ($\epsilon_{\rm f}=0.15$) was motivated in \S\,\ref{sec:agn_feedback}. In contrast to $\Delta T_{\rm SF}$, whose value is fixed by the need to suppress numerical radiative losses and to adequately sample the feedback process, the freedom to adjust $\Delta T_{\rm AGN}$, which determines the energetics and intermittency of AGN feedback events, can be motivated. We therefore explore simulations with different AGN heating temperatures but, in terms of the calibration of galaxy properties, $\Delta T_{\rm AGN}$ only (weakly) affects the stellar mass of galaxies with $M_\star \gtrsim 10^{11}\Msun$ \citep[it does, however, impact markedly upon the properties of the intragroup and intracluster media, e.g.][]{Le_Brun_et_al_14,Schaye_et_al_15_short}. For the purposes of reproducing the present-day GSMF, AGN feedback is primarily calibrated by varying $C_{\rm visc}$, which broadly governs the mass scale at which AGN feedback becomes efficient. Since this mass scale is weakly dependent upon $C_{\rm visc}$, models can adopt values of this parameter that differ by orders of magnitude.

The calibration of phenomenological components of galaxy formation models is a practice that was established during the development of the first generation of semi-analytic galaxy formation models\footnote{Modern semi-analytic models often adopt distribution sampling techniques to efficiently calibrate their many free parameters \citep{Kampakoglou_Trotta_and_Silk_08,Henriques_et_al_09,Bower_et_al_10,Lu_et_al_12,Mutch_Poole_and_Croton_13,Henriques_et_al_13,Henriques_et_al_14}.} \citep{Kauffmann_White_and_Guiderdoni_93,Cole_et_al_94,Somerville_and_Primack_99,Cole_et_al_00,Hatton_et_al_03}. Semi-analytic models are built upon the framework of dark matter halo merger trees, so the parameters they adopt for governing feedback must be coupled to simplified models for the structure of galaxies and their interstellar and circumgalactic media. In contrast, hydrodynamical simulations enable feedback properties to be specified, if so desired, based on the physical conditions \textit{local} to newly formed star particles. 

Thus far the capability to specify feedback parameters in this fashion has only been exploited by a limited number of groups, each using a very similar feedback implementation: kinetically-driven winds that are launched outside of the ISM (by decoupling the wind particles from hydrodynamic forces), and whose properties are imposed by specifying their initial velocity and mass loading factor ($\eta = \dot{M}_{\rm wind}/\dot{M}_\star$) as a function of the properties of the dark matter environment, for example the gravitational potential or velocity dispersion. Simulations adopting this implementation have been used to investigate the establishment of the GSMF \citep{Oppenheimer_et_al_10,Puchwein_and_Springel_13,Vogelsberger_et_al_13} and the formation of disc galaxies similar to the Milky Way \citep{Marinacci_Pakmor_and_Springel_14}, to examine the observed properties of some intergalactic absorption systems \citep{Oppenheimer_and_Dave_06,Oppenheimer_and_Dave_09,Vogelsberger_et_al_14Nat_short} and to reproduce the Local Group satellite population \citep{Okamoto_et_al_10}. 

This implementation is well-motivated since, by temporarily decoupling winds and specifying their properties as a function of dark matter properties, it affords simulations the best opportunity to achieve numerical convergence. However, it precludes the full exploitation of the hydrodynamics calculation. The physical properties of outflows are almost certainly dependent upon the local (baryonic) conditions of the ISM, and these properties are available to use as inputs to subgrid feedback models. Since the philosophy adopted for the EAGLE project is to calibrate the feedback scheme, the convergence demands placed upon the adopted subgrid models are relaxed, presenting the appealing opportunity to couple the value of subgrid parameters (e.g. $f_{\rm th}$) to the baryonic properties of the local ISM\footnote{S15 introduced the nomenclature ``weak convergence'' to describe the consistency of simulation outcomes in the case that subgrid parameters are recalibrated when the resolution is changed, as opposed to ``strong convergence'' in the case of holding the parameters fixed.}.

\subsection{Calibrating the star formation feedback efficiency}
\label{sec:calibrating_fth}

The role of $f_{\rm th}$ in shaping the $z=0.1$ GSMF is investigated in this section. Examination of the previous generation of simulations upon which the EAGLE project is based indicates that AGN feedback is a necessary ingredient for regulating the growth of massive galaxies \citep{Crain_et_al_09_short,Schaye_et_al_10_short,Haas_et_al_13a} and establishing the gas-phase properties of galaxy groups \citep{McCarthy_et_al_10_short,McCarthy_et_al_11,Le_Brun_et_al_14}. Four calibrated simulations are explored, each featuring energy feedback associated with both star formation and the growth of BHs. All four simulations adopt $\Delta T_{\rm AGN} = 10^{8.5}\K$, and each features a constant value of $C_{\rm visc}$ that is allowed to differ between simulations such that, when combined with the function specifying $f_{\rm th}$, the resulting $z = 0.1$ GSMF features a break close to the observed scale of $M^\star \sim 10^{10.5}\Msun$. 

In three of the models, asymptotic efficiencies of $f_{\rm th,max}=3$ and $f_{\rm th,min}=0.3$ are used. Values greater than unity can be motivated physically, since there are sources of energy feedback other than type II SNe, and indeed such sources are often invoked in simulations of galaxy formation, for example stellar winds and radiation pressure \citep{Stinson_et_al_13,Hopkins_et_al_14} or cosmic rays \citep{Jubelgas_et_al_08,Booth_et_al_13}. However the primary motivation for appealing to $f_{\rm th}>1$ here is to offset numerical losses that result from the finite resolution of the simulations. There are two means by which finite resolution introduces
artificial losses. The first, as discussed in \S\ref{sec:sf_feedback} (equation \ref{eq:nhtc}), being
that there exists a maximum density above which stochastic thermal
feedback is inefficient, because the pressure gradient established
by feedback is erased by radiative cooling before it is able to exert
mechanical work on the gas. The second stems from the inability
of large cosmological simulations to model the formation of the
earliest generation of stars. As discussed by \citet{Haas_et_al_13a},
this means that the first generation of galaxies that form in
simulations do so within haloes that have not been subject to
feedback, and hence exhibit unrealistically high gas fractions and
star formation efficiencies. Appealing to $f_{\rm th}>1$ for these galaxies
partially compensates for this unavoidable shortcoming.

\subsubsection*{FBconst}
The simplest model injects into the ISM a fixed quantity of energy per unit stellar mass formed, independent of local conditions. This value corresponds to the total energy liberated by type II SNe, i.e. $f_{\rm th}=1$. The adopted subgrid viscosity parameter for BH accretion is $C_{\rm visc} = 2\pi \times 10^3$. Although the injected energy is independent of local conditions, scale-dependent macroscopic radiative losses can nonetheless develop self-consistently, for example due to differences in the metallicity (and hence cooling rate) of outflowing gas, the ram pressure at the disc-CGM interface, or the depth of the potential well. This model therefore provides a baseline against which it is possible to assess the degree to which the overall physical losses need to be established by calibrating losses on subgrid scales. Because $f_{\rm th}=1$ represents the uncalibrated case, there is no reason to expect that FBconst will reproduce the observed $z=0.1$ GSMF. However, we will see later that it does do so, but fails to reproduce the observed sizes of disc galaxies.

\begin{figure}
\includegraphics[width=\columnwidth]{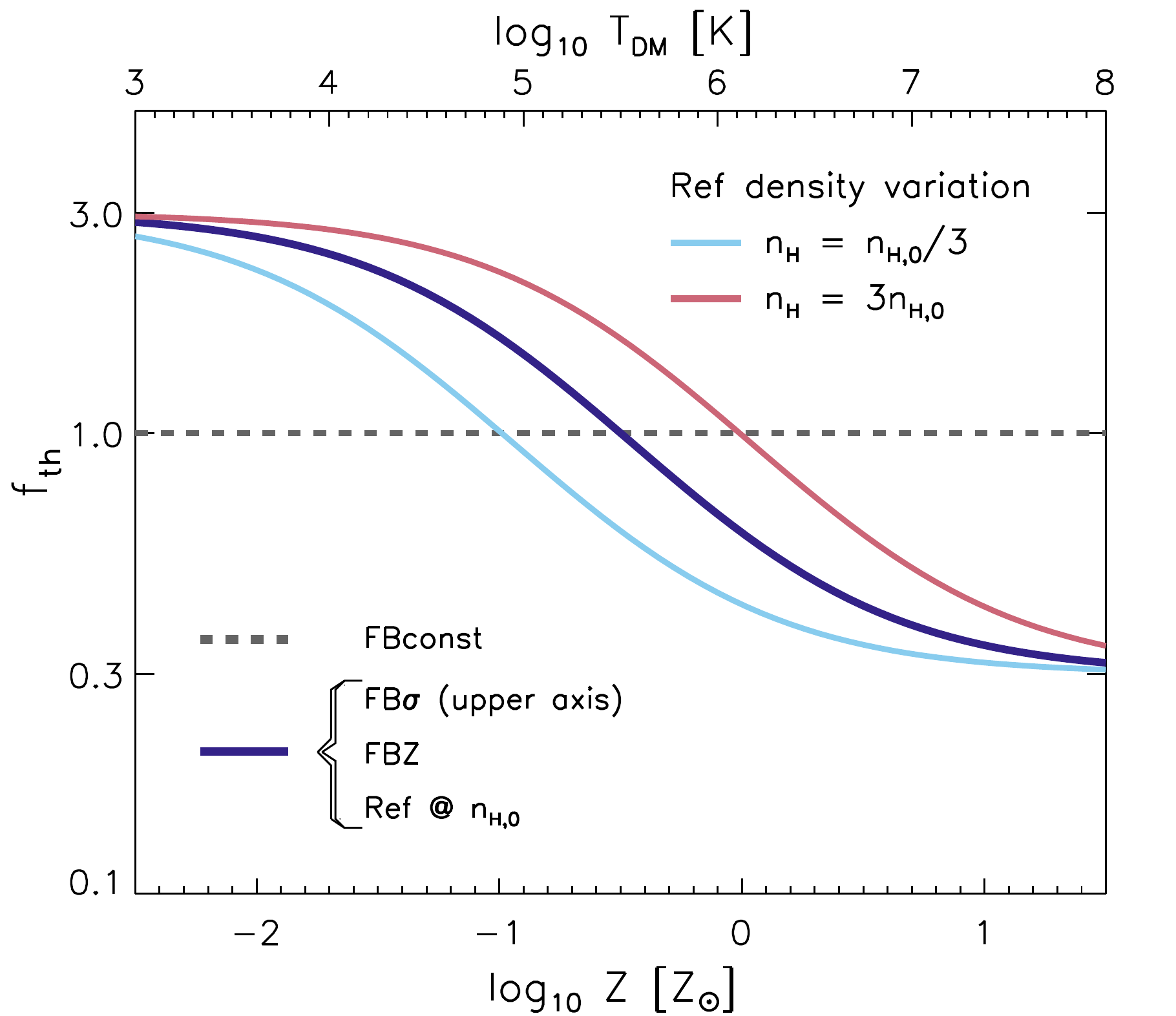}
\caption{The fraction of the energy budget due to type II SNe feedback that is used for thermal heating, $f_{\rm th}$, in the FBconst, FB$\sigma$, FBZ and Ref models. The FBconst model, represented by the dashed grey line, adopts $f_{\rm th}=1$, independent of local conditions. $f_{\rm th}$ declines smoothly as a function of $T_{\rm DM}$ in the FB$\sigma$ model (equation \ref{eq:fth(T)}), and as a function of $Z$ in the FBZ model (equation \ref{eq:fth(Z)}). The upper axis is aligned and scaled such that both FB$\sigma$ (upper axis) and FBZ (lower axis) are described by the dark blue curve (no physical correspondence between $T_{\rm DM}$ and $Z$ is implied by this alignment). The Ref model adds a density dependence to FBZ (equation \ref{eq:fth(Z,n)}), such that for stars forming from gas with $n_{\rm H}<n_{\rm H,0}$ the $f_{\rm th}$ function is shifted to lower values (e.g. cyan curve for $n_{\rm H}=n_{\rm H,0}/3$) and vice versa (e.g. red curve for $n_{\rm H}=3n_{\rm H,0}$).}
\label{fig:fth_curve}
\end{figure}

\subsubsection*{FB$\sigma$}

This model adopts the popular convention of prescribing feedback parameters according to local conditions inferred from neighbouring dark matter particles \citep{Oppenheimer_and_Dave_06,Oppenheimer_and_Dave_08,Okamoto_et_al_10,Oppenheimer_et_al_10,Puchwein_and_Springel_13,Vogelsberger_et_al_13,Khandai_et_al_14,Vogelsberger_et_al_14_short}. However, because these studies all adopt kinetically-driven outflows imposed outside of the ISM, and specify the properties of the outflows by imposing a mass loading and initial wind velocity, our implementation is significantly different. Rather than specifying the properties of the outflows, similar behaviour on macroscopic scales is sought as an \textit{outcome} of the stochastic heating implementation, by calibrating the efficiency, $f_{\rm th}$, as a function of $\sigma_{\rm DM}^2$. The latter is the square of the 3-dimensional velocity dispersion of dark matter particles within the smoothing kernel of a star particle at the instant it is born, and is a proxy for the characteristic virial scale of the star particle's environment, 
\begin{equation}
T_{\rm DM}= \frac{\mu m_{\rm p} \sigma_{\rm DM}^2}{3k} \simeq (4\times 10^5\K) \mu \left( \frac{\sigma_{\rm DM}}{100\,{\rm km s^{-1}}} \right)^2. 
\end{equation}
For simplicity, we assume at all times the mean molecular weight of a fully ionized gas with primordial composition, $\mu = 0.591$. The fit to the $z=0.1$ GSMF for this model is achieved with a slightly higher subgrid viscosity for BH accretion than is the case for FBconst, $C_{\rm visc} = 2\pi \times 10^2$.

Since the properties of star formation-driven outflows are linked to the state of local dark matter only via gravitational forces, no physical motivation for $f_{\rm th}(T_{\rm DM})$ is sought. The adopted functional form simply maximises the feedback efficiency (by minimising putative subgrid radiative losses) in low-mass galaxies, whilst reducing the feedback efficiency in more massive counterparts, where the conversion of gas into stars is known to be most efficient \citep[e.g.][]{Eke_et_al_05,Behroozi_Wechsler_and_Conroy_13,Moster_Naab_and_White_13}. At higher masses still, AGN feedback is assumed to dominate the regulation of star formation, so a low star formation feedback efficiency for high-dispersion environments is reasonable. The adopted functional form of $f_{\rm th}$ is a logistic (sigmoid) function of $\log_{10} T_{\rm DM}$,
\begin{equation}
f_{\rm th} = f_{\rm th,min} + \frac{f_{\rm th,max} - f_{\rm th,min}}
{1 + \left (\frac{T_{\rm DM}}{10^5\K}\right )^{n_{\rm T}}}, 
\label{eq:fth(T)}
\end{equation}
shown in Figure\,\ref{fig:fth_curve} (dark blue curve, corresponding to the upper $x$-axis). The function asymptotes to $f_{\rm th,max}$ and $f_{\rm th,min}$ in the limits $T_{\rm DM} \ll 10^5\K$ and $T_{\rm DM} \gg 10^5\K$, respectively, and varies smoothly between these limits about $T_{\rm DM} = 10^5\K$ (or $\sigma_{\rm DM} \simeq 65\kms$). The parameter $n_{\rm T} > 0$ controls how rapidly $f_{\rm th}$ varies as the dark matter ``temperature'' scale deviates from $10^5\K$. The rather unnatural value $n_{\rm T} = 2/\ln{10}\simeq0.87$ follows from an early implementation of the functional form adopted in the feedback routine; an exponent of unity would yield similar results. 

\subsubsection*{FBZ}
Adjusting the subgrid radiative losses with the metallicity of the ISM assigns a physical basis to the functional form of $f_{\rm th}$. Physical losses associated with star formation feedback\footnote{A metallicity dependence for 
$\epsilon_{\rm f}$ (\S\,\ref{sec:agn_feedback}) is not explored, because metals are not expected to dominate the radiative losses at the higher temperatures associated with AGN feedback.} are likely to be more significant when the metallicity is sufficient for cooling from metal lines to dominate over the contribution from H and He. For temperatures $10^5\K < T < 10^7\K$, characteristic of outflowing gas in the simulations, the transition is expected to occur at $Z\sim0.1\Zsun$ \citep{Wiersma_et_al_09}. This qualitative behaviour is captured by the same functional form as equation \ref{eq:fth(T)}, replacing ($T_{\rm DM}$,$n_{\rm T}$,$10^5\K$) with ($Z$,$n_{\rm Z}$,$0.1\Zsun$) to obtain,
\begin{equation}
f_{\rm th} = f_{\rm th,min} + \frac{f_{\rm th,max} - f_{\rm th,min}}
{1 + \left (\frac{Z}{0.1\Zsun}\right )^{n_Z}},
\label{eq:fth(Z)}
\end{equation}
where $\Zsun = 0.0127$ is the solar metallicity \citep{AllendePrieto_Lambert_and_Asplund_01} and $n_{\rm Z} = n_{\rm T} = 2/\ln{10}$. This function corresponds to the dark blue curve and the lower $x$-axis in Figure\,\ref{fig:fth_curve}, with $f_{\rm th}$ asymptoting to $f_{\rm th,max}$ and $f_{\rm th,min}$ in the limits $Z \ll 0.1\Zsun$ and $Z \gg 0.1\Zsun$, respectively. Since galaxies tend to follow a tight relation between their mass and metallicity, the feedback efficiency is, as in FB$\sigma$, greatest for low-mass galaxies. Moreover, because metallicity characteristically decreases with redshift at fixed stellar mass, the feedback efficiency is weighted towards early cosmic epochs. This helps to partially decouple the growth of galaxies from the growth of their parent halo, which is thought to be a necessary condition for reproducing the observed number density evolution of low-mass galaxies in a CDM cosmogony \citep[e.g.][]{Weinmann_et_al_12,Henriques_et_al_13,Mitchell_et_al_14,White_Somerville_and_Ferguson_14}. The subgrid viscosity parameter for AGN is the same as per FB$\sigma$, $C_{\rm visc} = 2\pi \times 10^2$.

\begin{figure*}
\includegraphics[width=\columnwidth]{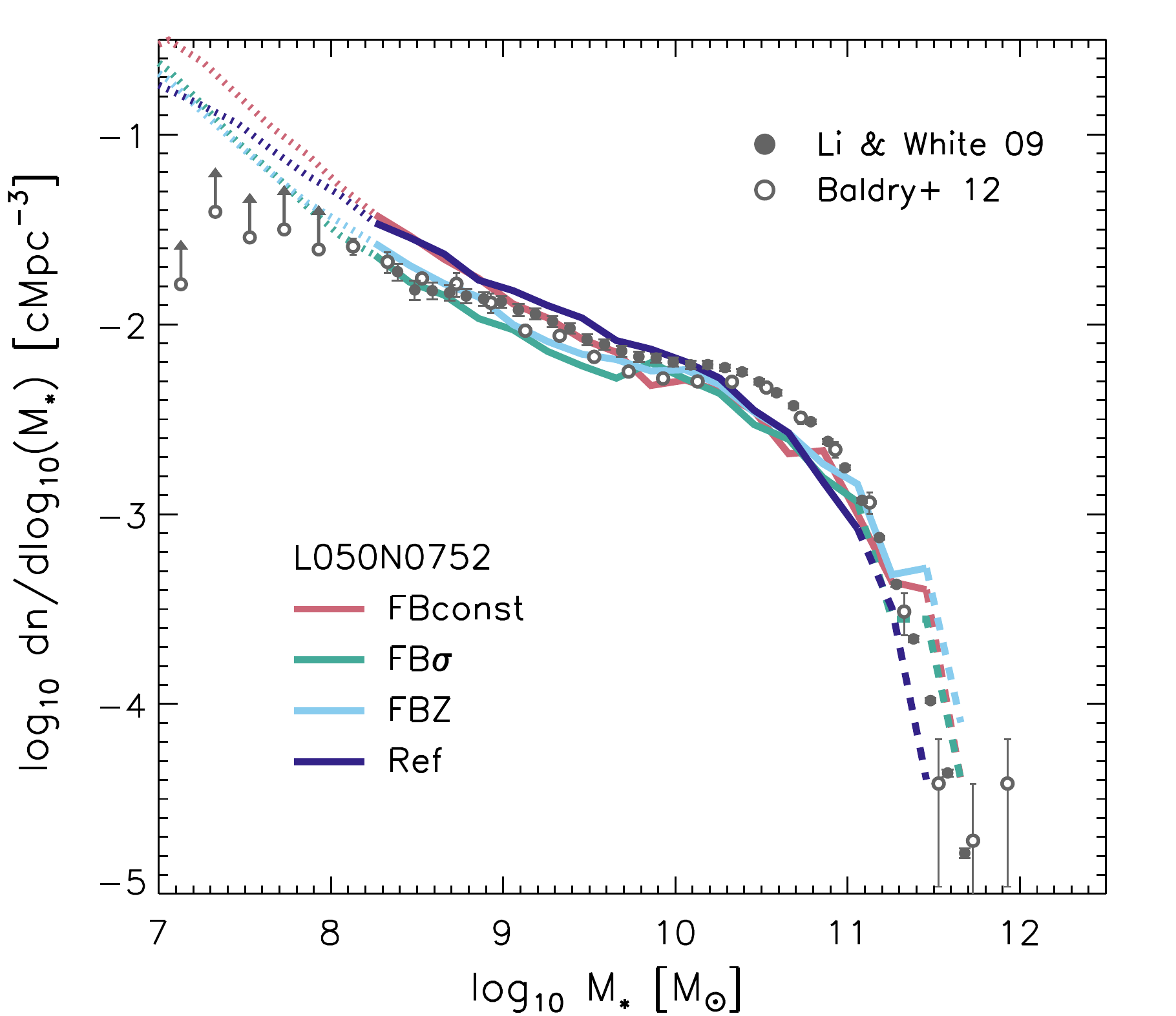}
\includegraphics[width=\columnwidth]{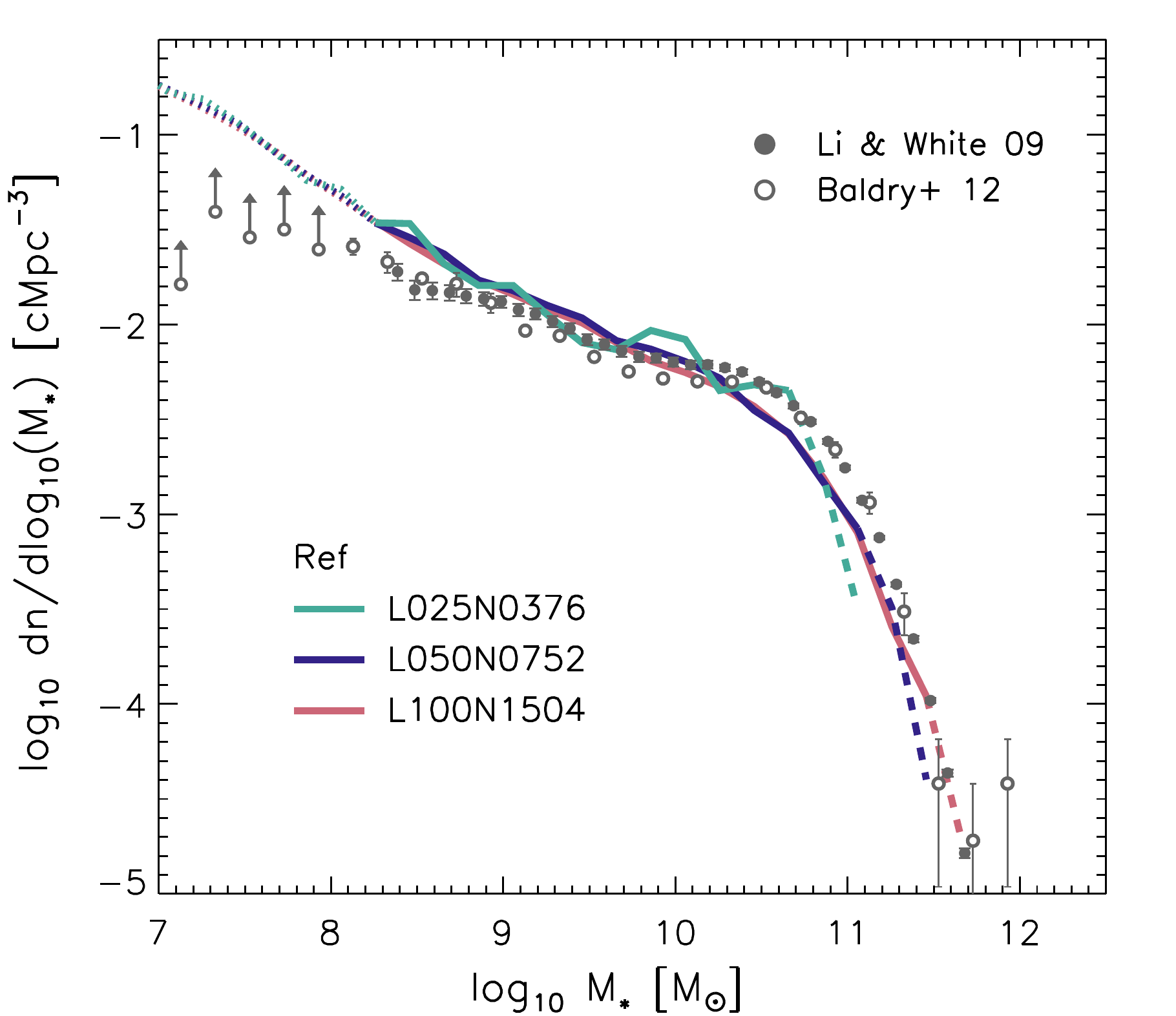}
\caption{\textit{Left:} The $z=0.1$ GSMF of the calibrated L050N0752 simulations, FBconst (red), FB$\sigma$ (green), FBZ (cyan) and Ref (dark blue). Curves are drawn with dotted lines where galaxies are comprised of fewer than 100 star particles, and dashed lines where the GSMF is sampled by fewer than 10 galaxies per bin. Data points show measurements with $1\sigma$ error bars from the SDSS \citep[][filled circles]{Li_and_White_09} and GAMA \citep[][open circles]{Baldry_et_al_12_short} surveys. The simulations each reproduce the observed number density of galaxies at fixed mass to $< 0.31\,{\rm dex}$, a precision unprecedented for hydrodynamical simulations. \textit{Right:} To illustrate convergence as the simulation volume is varied, the Ref model at intermediate resolution is shown in volumes of $L=25$, $50$ and $100\cMpc$. The Ref-L050N0752 (dark blue) and Ref-L100N1504 (red) GSMFs are consistent for $M_\star \lesssim 10^{11.5}\Msun$. The GSMF of Ref-L025N0376 is consistent with that of its larger counterparts for $M_\star < 10^{9.5}\Msun$, but samples large-scale structures poorly owing to its small volume, imprinting ``wiggles'' on the GSMF.}
\label{fig:mf_calibrated}
\end{figure*}

\subsubsection*{Ref (equivalently, FBZ$\rho$)}
A significant fraction of the star particles in the FB$\sigma$ and FBZ models form at densities greater than $\nHtc$, the critical density above which feedback energy is quickly radiated away (equation \ref{eq:nhtc}). The feedback associated with these high-density star formation events is therefore numerically inefficient. The consequences of this overcooling are explored later in \S\,\ref{sec:results_calibrated}. The overestimated losses can be compensated by introducing a density dependence in the expression for $f_{\rm th}$:
\begin{equation}
f_{\rm th} = f_{\rm th,min} + \frac{f_{\rm th,max} - f_{\rm th,min}}
{1 + \left (\frac{Z}{0.1\Zsun}\right )^{n_Z} \left (\frac{n_{\rm H,birth}}{n_{{\rm H},0}}\right )^{-n_n}},
\label{eq:fth(Z,n)}
\end{equation}
where $n_{\rm H,birth}$ is the density of a gas particle at the instant it is converted into a star particle. The feedback efficiency therefore increases with density at fixed metallicity, whilst respecting the original asymptotic values. The choice of $n_{\rm H,0}=0.67\,\cmcubed$ was guided by a suite of small test simulations, which also indicated that the adoption of $n_{\rm n}=n_{\rm Z}$ is sufficient to reproduce the $z=0.1$ GSMF. The effect of the additional density term is illustrated in Figure \ref{fig:fth_curve}, where it can be seen that for $n_{\rm H}=n_{\rm H,0}$ the functional form adopted by Ref is identical to that of FBZ, but the $f_{\rm th}$ curve is shifted to lower (higher) values for stars forming from lower (higher) density gas. Only the shift to higher $f_{\rm th}$ for higher density gas (at fixed metallicity) is required to offset numerical losses, but for simplicity we choose to include the shift to lower efficiency at low density that is implied by the adopted function. Such a density dependence may also have a physical basis: because the star formation law has a supra-linear dependence on surface density, the feedback energy injection rate per unit volume increases with density. At fixed density, a higher energy injection rate corresponds to higher temperatures and longer cooling times. Therefore, feedback associated with clustered star formation is expected to lead to lower radiative losses \citep[e.g.][]{Heiles_90,Creasey_Theuns_and_Bower_13,Krause_et_al_13,Nath_and_Shchekinov_13,Roy_et_al_13,Keller_et_al_14}, and vice-versa. The density-dependent feedback efficiency adopted here (equation \ref{eq:fth(Z,n)}) ensures that, when integrating over all star formation events in the simulation, the mean and median values of $f_{\rm th}$ remain close to unity (they are 1.06 and 0.70, respectively, for Ref-L0100N1504). The subgrid viscosity parameter for BH accretion is lower in Ref than in the other calibrated models, $C_{\rm visc} = 2\pi \times 10^0$.

\section{Results}
\label{sec:results}

This section begins with an examination of the calibrated simulations. In \S\ref{sec:reference_variations}, simulations featuring single parameter variations to the reference model are briefly introduced, and the impact of the changes on the resulting galaxy population are explored.

\subsection{Examination of the calibrated simulations}
\label{sec:results_calibrated}

We begin by examining the models calibrated to reproduce the $z=0.1$ GSMF. Besides the GSMF, the evolution of the comoving stellar mass density of each simulation is examined, as are the sizes and specific star formation rates (SSFRs) of galaxies at $z=0.1$. As discussed by S15, the GSMF and galaxy sizes were used for the calibration process, so are not presented as predictions. Although the model parameters were not adjusted in order to improve the correspondence between the simulated and observed SSFRs, the former were inspected throughout the calibration process, and hence the predictions of the SSFRs are not ``blind''. The aim of this exercise is to examine how different implementations of physical processes impact upon the properties of galaxies and their environments. As part of this procedure, the conditions of the ISM from which stars are born are also explored.

\subsubsection{The galaxy stellar mass function}
\label{sec:gsmf_calibrated}

The $z=0.1$ GSMFs of the four calibrated simulations, FBconst (red curve), FB$\sigma$ (green), FBZ (cyan) and Ref (dark blue), run using L050N0752 initial conditions, are shown in the left-hand panel of Figure \ref{fig:mf_calibrated}. The right-hand panel also shows the GSMF of the reference model at intermediate resolution in a $L=100\cMpc$ volume (Ref-L100N1504, red curve), which was introduced by S15, and in a smaller $L=25\cMpc$ realisation (Ref-L025N376, green), which is used in \S\ref{sec:reference_variations} as a baseline against which to compare several single parameter variations of the reference model. Maintaining the convention established by S15, curves are drawn with dotted lines where galaxies are less massive than 100 (initial mass, $m_{\rm g}$) baryonic particles, as resolution tests (presented in S15) indicate that sampling errors due to finite resolution are significant in this regime. At the high-mass end, curves are drawn with dashed lines where the GSMF is sampled by fewer than 10 galaxies per ($\Delta \log_{10}{M}_* = 0.2$) bin. Data points represent the GSMF inferred from the Sloan Digital Sky Survey \citep[][SDSS, filled circles]{Li_and_White_09} and the Galaxy and Mass Assembly \citep[][GAMA, open circles]{Baldry_et_al_12_short} survey. In both cases, the data have been adjusted for consistency with the value of the Hubble parameter adopted by the simulations. GAMA measurements on scales $M_\star < 10^8\Msun$ are drawn as lower limits, since the data are expected to be incomplete at the typical surface brightness associated with these galaxies \citep{Baldry_et_al_12_short}.

In the range of stellar masses where the mass resolution and volume of the simulation enable a robust measurement of the GSMF ($10^8 < M_\star \lesssim 10^{11} \Msun$), the number density of galaxies at fixed stellar mass produced by each of the four calibrated models is consistent with the observational data to $< 0.31\,{\rm dex}$. This precision is comparable to the systematic uncertainty associated with spectrophotometric techniques for inferring galaxy stellar masses, indicating that a more precise reproduction of the GSMF may be unwarranted\footnote{The GSMF is also impacted upon by other systematic effects, such as completeness and extinction corrections, and background subtraction.} \citep[e.g.][]{Conroy_Gunn_and_White_09,Pforr_Maraston_and_Tonini_12}. This degree of consistency with observational data is typical of that associated with semi-analytic galaxy formation models. The reproduction of $z=0.1$ GSMF by multiple cosmological hydrodynamical simulations featuring feedback efficiencies governed by such distinct schemes is unprecedented in the literature (see also Fig. 5 of S15). 

The detailed confrontation of the EAGLE reference model with observational data presented by S15 was based on the Ref-L100N1504 simulation. Since running multiple L100N1504 simulations is, at present, computationally prohibitive, the calibrated models have been run with L050N0752 initial conditions. It is therefore necessary to confirm that the Ref-L050N0752 simulation yields a GSMF that is consistent with that of its $L=100\cMpc$ counterpart. Comparison of the Ref-L050N0752 and Ref-L100N1504 curves in the right-hand panel of Figure \ref{fig:mf_calibrated} indicates that the GSMFs of these simulations are consistent to a high precision ($<0.063\,{\rm dex}$ at fixed stellar mass) over the range for which both simulations are well sampled ($M_\star \lesssim 10^{11.5}\Msun$). An $L=50\cMpc$ volume is therefore sufficient to capture the effects of subgrid physics on all but the most massive galaxies seen in observational surveys of the local Universe. The Ref-L025N0376 simulation tracks its larger counterparts on scales of $M_\star < 10^{9.5}\Msun$, but lacks the volume required to sample more massive scales precisely, as is clear from the ``wiggles'' imprinted on the GSMF. 

We do not explore resolution convergence here; S15 demonstrated the strong convergence behaviour of the $z=0.1$ GSMF for the reference model, and established that recalibration of the subgrid parameters enables competitive and well-understood weak convergence behaviour for a broad range of observational diagnostics. 

\subsubsection{Galaxy sizes}
\label{sec:sizes_calibrated}

\begin{figure}
\includegraphics[width=\columnwidth]{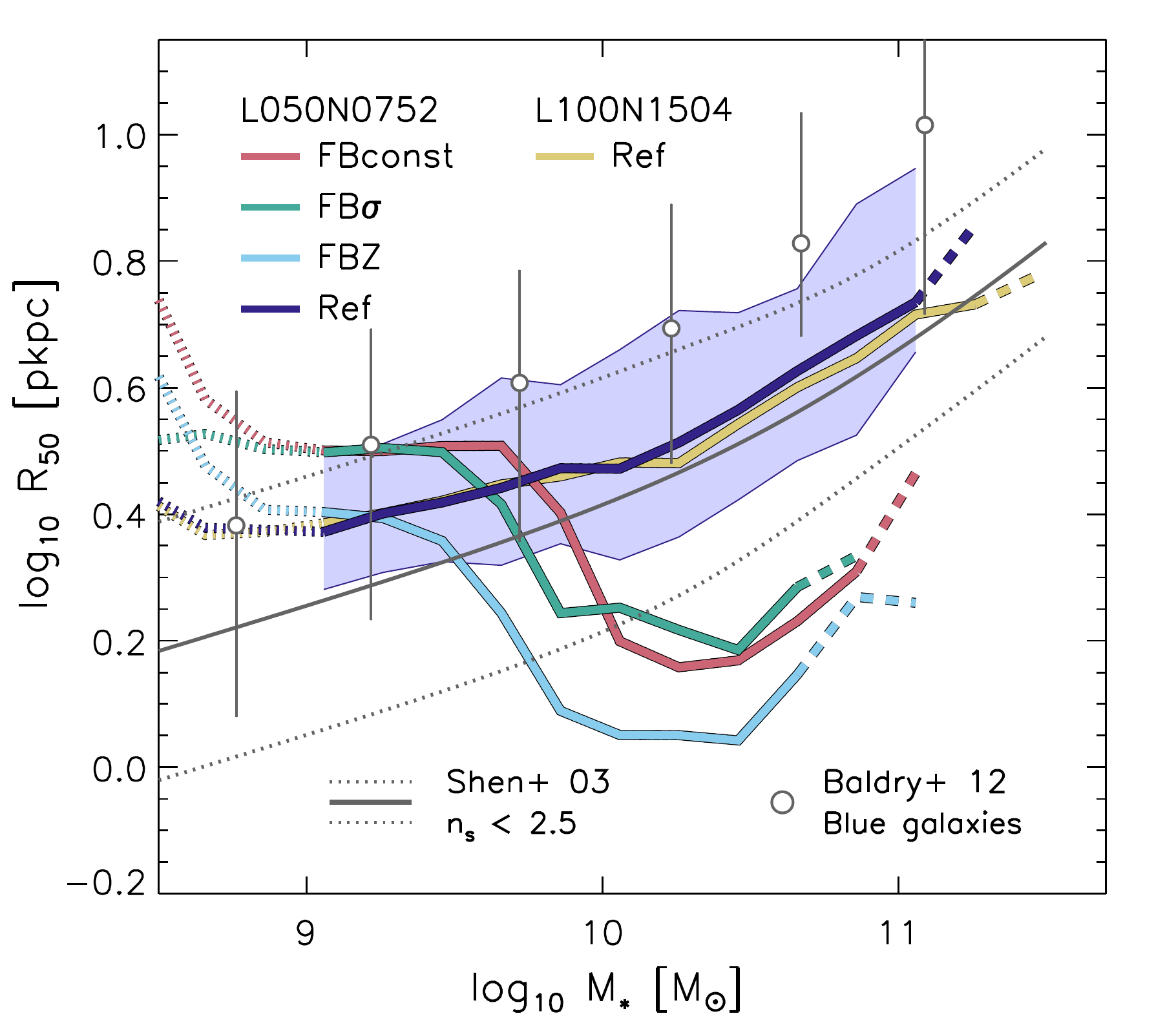}
\caption{The sizes, at $z=0.1$, of disc galaxies in the four simulations that were calibrated to reproduce the present-day GSMF. Sizes from the Ref-L100N1504 (presented by S15) are also shown (yellow curve), to demonstrate consistency between volumes. Size, $R_{50}$, is defined as the half-mass radius (in proper kpc) of a S\'{e}rsic profile fit to the projected, azimuthally-averaged stellar surface density profile of a galaxy, and those with S\'{e}rsic index $n_{\rm s}<2.5$ are considered disc galaxies. Curves show the binned median sizes, and are drawn with dotted lines below a mass scale of 600$m_{\rm g}$, and a dashed linestyle where sampled by fewer than 10 galaxies per bin. The $1\sigma$ scatter about the median of Ref is denoted by the blue shaded region. The solid and dotted grey lines show the median and $1\sigma$ scatter of sizes for $n_{\rm s}<2.5$ galaxies inferred from SDSS data by \citet{Shen_et_al_03}, whilst grey data points and error bars show sizes of blue galaxies inferred by \citet{Baldry_et_al_12_short} from GAMA data. Only the Ref model successfully reproduces the observed GSMF and galaxy sizes at $z=0.1$.}
\label{fig:sizes_calibrated}
\end{figure}

Following \citet{McCarthy_et_al_12}, we characterise the morphology of galaxies by fitting S\'{e}rsic profiles to their projected, azimuthally-averaged surface density profiles. The size of a galaxy is then equated to its effective radius, $R_{50}$, the radius enclosing $50$ percent of the stellar mass when the profile is integrated to infinity. The scaling of this quantity with stellar mass for disc galaxies is shown in Figure \ref{fig:sizes_calibrated}. As in \citet{Shen_et_al_03}, whose size measurements from SDSS data are overplotted, disc galaxies are defined to be those with S\'{e}rsic indices $n_{\rm s}<2.5$. The binned median is plotted for each calibrated simulation. Dashed lines are used where the median is sampled by fewer than 10 galaxies, and a dotted linestyle is used for bins corresponding to stellar masses $ < 600 m_{\rm g}$; this scale was shown by S15 to be the minimum for which size measurements are robust. In the regime between the limits of sufficient resolution and adequate galaxy sampling, the $1\sigma$ (i.e. $16^{\rm th}$ to $84^{\rm th}$ percentile) scatter about the median of Ref is shown as a blue shaded region. 

\begin{figure*}
\includegraphics[width=\textwidth]{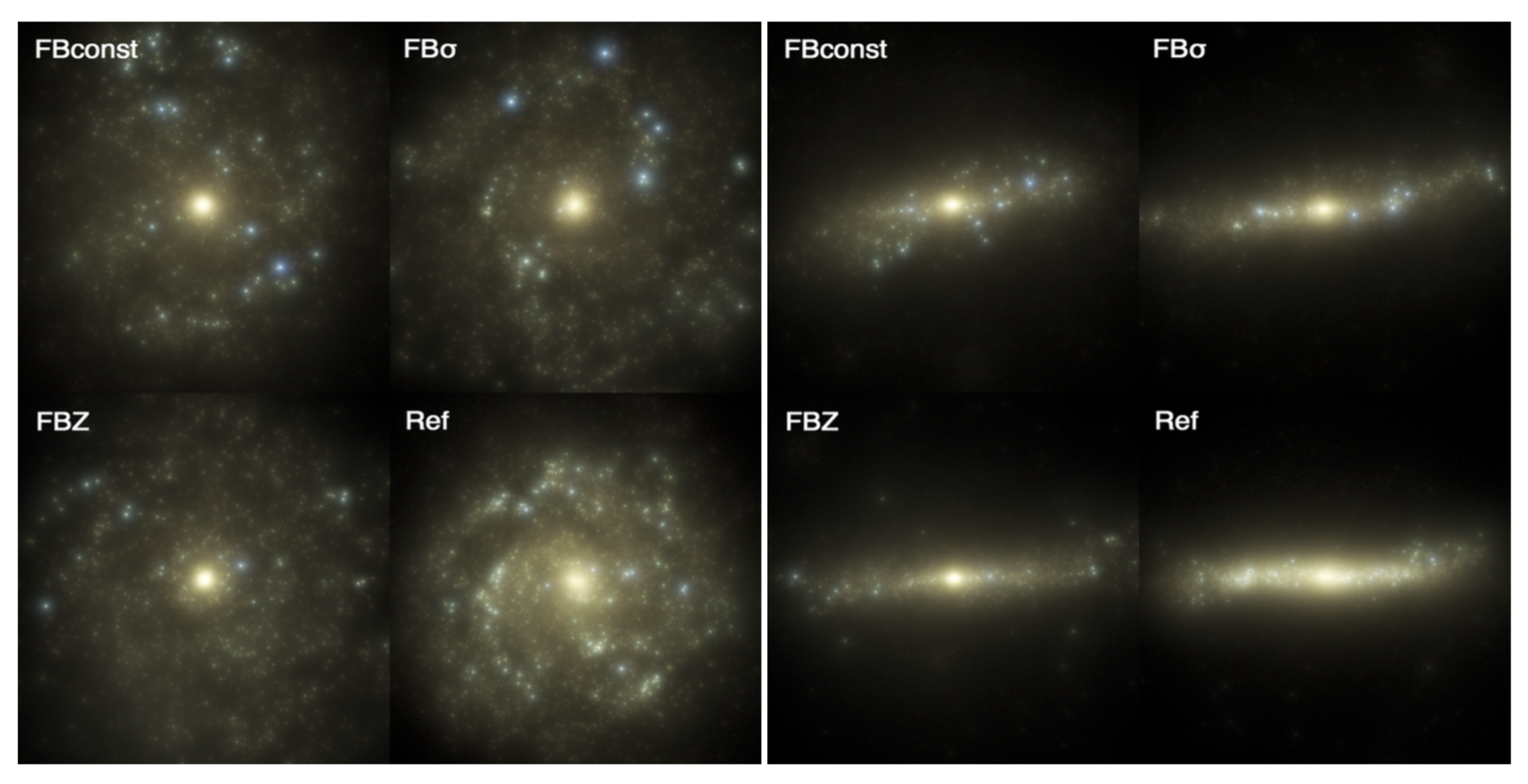}
\caption{Face-on (\textit{left}) and edge-on (\textit{right}) projections, at $z=0.1$, of the galaxy formed within the same $M_{200} \sim 10^{12}\Msun$ halo in the four simulations that were calibrated to reproduce the present-day GSMF. Each image subtends a field of view of $100\,{\rm pkpc}$, and is a composite of SDSS $u$-, $g$- and $r$-band emission maps generated with the radiative transfer software \textsc{skirt}. The overall size of the optical envelope of the galaxy is similar in each case, but the distribution and dynamics of the stars differ markedly between Ref and the other models. In the latter, the galaxy forms an unrealistically compact bulge component at early epochs and exhibits too little ongoing star formation (blue-coloured concentrations) in the extended disc. Consequently, only in Ref does the galaxy exhibit an effective radius, $R_{50}$, that is consistent with the observed size-mass relation for disc galaxies.}
\label{fig:3band}
\end{figure*}

The reference model tracks the observed relation closely, with the medians of the simulation and the observational measurements being offset by $0.1-0.2~{\rm dex}$. This offset is comparable to the systematic offset between the median of those data and the median sizes measured by \citet{Baldry_et_al_12_short} based on GAMA observations of blue galaxies (defined using an $r$-band magnitude-dependent colour threshold). Also shown in Figure \ref{fig:sizes_calibrated} are the size measurements presented by S15 for the larger Ref-L100N1504 volume (yellow curve) demonstrating that galaxy sizes are unaffected by the volume of the simulation, as expected. In contrast to Ref, the FBconst, FB$\sigma$ and FBZ models are inconsistent with the observed size-mass relation. Galaxies with $M_\star \gtrsim 10^{10}\Msun$ cease to follow the observed relation between size and mass, and become much too compact, with median sizes only a few times the gravitational softening scale ($\epsilon_{\rm prop} = 0.7\,{\rm pkpc}$). For this reason, these models are not considered satisfactory when applying the EAGLE calibration criteria. We note that the measurements of both \citet{Shen_et_al_03} and \citet{Baldry_et_al_12_short} are based on $r$-band photometry, and that isophotal radii are sensitive to the band in which they are measured. \citet{Lange_et_al_15_short} recently exploited the multiband photometry of the GAMA survey to assess the magnitude of this sensitivity, concluding that the observed size of disc galaxies of mass $M_\star = 10^{10}\Msun$ is typically only $\sim10$ percent smaller in the $K_{\rm s}$-band (the best proxy for the true stellar mass distribution) than in the $r$-band. We can therefore be confident that our conclusions here are unlikely to be strongly affected by, for example, attenuation by interstellar dust. Detailed tests of mock observations derived by coupling EAGLE to the \textsc{skirt} radiative transfer algorithm \citep{Baes_et_al_11} will be presented in a forthcoming study (Trayford et al. \textit{in prep}).

The calibrated simulations adopt identical initial conditions, enabling galaxies to be compared individually as well as statistically. In Figure \ref{fig:3band}, the galaxy that forms within the same dark matter halo is shown for each of the calibrated simulations at $z=0.1$. The halo was selected at random from those with mass in the Ref simulation $8 \times 10^{11} \Msun < M_{200} < 2 \times 10^{12} \Msun$. In each simulation, it therefore hosts a galaxy whose stellar mass ($M_\star \sim 2 \times 10^{10}\Msun$) corresponds to the minimum of the size-mass relation exhibited by the FBconst, FB$\sigma$ and FBZ simulations. Each image subtends a field of view $100\,{\rm pkpc}$ on a side, and is a composite comprised of monochromatic SDSS $u$-, $g$-, and $r$-band emission maps. The maps were generated with \textsc{skirt} \citep{Baes_et_al_11}, which considers the photometric properties of the stellar populations and the estimated dust distribution, the latter being inferred from the predicted metallicity of the ISM. The same mapping between physical flux and pixel luminosity is adopted in each panel. The galaxy is shown face-on (\textit{left-hand panels}) and edge-on (\textit{right-hand panels}), oriented about the angular momentum vector of the star particles comprising the galaxy. 

Visual inspection indicates that the outer envelope of the galaxy, corresponding to an $r$-band surface brightness of $\mu_{\rm r} \sim 28~{\rm mag~arcsec}^{-2}$, is similar in each simulation. However, the distribution of stars within that radius differs markedly between Ref and the other models, and this strongly influences the effective radius. In the FBconst, FB$\sigma$ and FBZ simulations, the galaxy forms a massive, compact bulge that dominates the overall stellar distribution. In the Ref simulation the star-forming disc component, seen clearly in the face-on images as blue-coloured concentrations distributed over all radii, comprises a greater fraction of the mass. Based on a dynamical decomposition similar to the orbital circularity method of \citet{Abadi_et_al_03a}, the bulge-to-total ratio of the galaxy in the FBconst, FB$\sigma$, FBZ and Ref simulations is 0.47, 0.43, 0.50 and 0.30, respectively. 

Many studies in the literature conclude that the ability of a hydrodynamical simulation to reproduce, approximately, the low-redshift GSMF (or, in the case of `zoom' simulations, the $M_\star - M_{\rm halo}$ relation as inferred by e.g. subhalo abundance matching) to be a metric of success, without comparing to the observed size of galaxies \citep[e.g.][]{Okamoto_et_al_10,Oppenheimer_et_al_10,Munshi_et_al_13_short,Puchwein_and_Springel_13,Vogelsberger_et_al_13}. However the reproduction here of the observed $z=0.1$ GSMF with a number of models that yield unrealistically compact galaxies, highlights the importance of simultaneously calibrating models with observational diagnostics that are complementary to the GSMF. In Sections \ref{sec:rhostar_calibrated}, \ref{sec:ssfr_calibrated} and \ref{sec:birth_conditions_calibrated} we turn to the examination of diagnostics that were not considered during the calibration process. It is shown that models that yield unrealistically compact galaxies also fail to reproduce the observed star formation history and present-day specific star formation rate of the galaxy population. We demonstrate that the formation of compact galaxies is a consequence of numerical radiative losses becoming severe in high density gas, thus artificially suppressing the efficiency of energy feedback.

\subsubsection{Comoving stellar mass density}
\label{sec:rhostar_calibrated}

\begin{figure}
\includegraphics[width=\columnwidth]{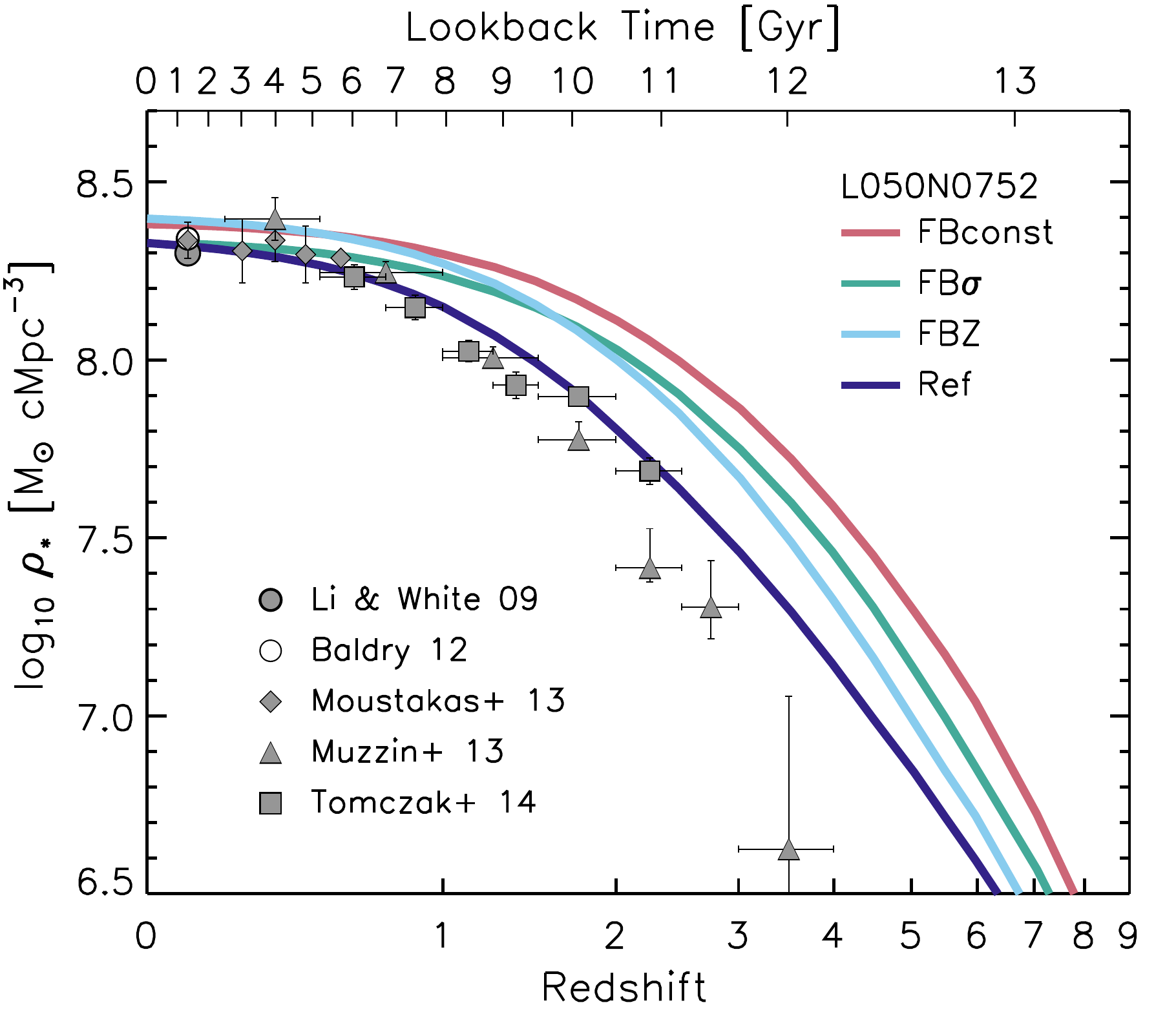}
\caption{The evolution of the comoving stellar mass density of the calibrated EAGLE models. Data points correspond to measurements from several observational surveys spanning $0 < z < 4$ (see text for details). Although the models each broadly reproduce the observed GSMF at $z=0.1$, they exhibit markedly different star formation histories, and the FBconst model in particular forms too much stellar mass at early times. Taking the observations at face value, only the Ref model is broadly consistent with the observational constraints for $z\lesssim 2$.}
\label{fig:rhostar_calibrated}
\end{figure}

By construction, the calibrated simulations yield similar volumetric stellar mass densities at $z = 0.1$. The evolution of stellar mass density in the four simulations can differ, however, because the history of energy injection from feedback varies between the models. Figure \ref{fig:rhostar_calibrated} shows the evolution of the comoving, instantaneous\footnote{Stellar evolution mass loss by star particles is accounted for.} stellar mass density in the four calibrated L050N0752 simulations. The total comoving density of stars in each simulation is shown; in a companion paper, \citet{Furlong_et_al_14_short} excluded diffuse intracluster light (by considering only those stars within $30\pkpc$ of galactic centres) to mimick observational measurements, and recovered densities in the Ref-L100N1504 simulation that were lower by approximately 20 percent for $z \lesssim 1$.

Data points represent the comoving stellar mass density inferred from a number of complementary observational analyses. Where necessary, the data have been adjusted to adopt the same IMF and Hubble parameter as the simulations. The filled and open circles at $z \sim 0.1$ represent the integration of the SDSS \citep{Li_and_White_09} and GAMA \citep{Baldry_et_al_12_short} GSMFs shown in Figure \ref{fig:mf_calibrated}, respectively. Diamonds represent measurements over the redshift interval $0.1 < z < 0.9$ inferred from a combined sample of SDSS and PRIMUS data presented by \citet{Moustakas_et_al_13_short}, triangles represent measurements over the redshift interval $0.2 < z < 4$ inferred from UltraVISTA data by \citet{Muzzin_et_al_13_short}, and squares represent measurements over the redshift interval $0.625 < z < 2.25$ inferred from ZFOURGE data by \citet{Tomczak_et_al_14_short}. Data from surveys that overlap in redshift interval are shown in order to illustrate, broadly, the degree of systematic uncertainty and field-to-field variance in the measurements.

The stellar mass densities of the four simulations are consistent to $\lesssim 0.1\,{\rm dex}$ at $z=0.1$. The FBconst simulation, however, forms stars too rapidly at early epochs. The stellar mass density inferred from observations at $z \sim 1$ is in place in this simulation prior to $z=2$, whilst the evolution at intermediate redshifts ($1\lesssim z \lesssim 3$) is, by necessity, then too weak. The models that allow the star formation feedback efficiency to vary as a function of the local environment track the observed build up of stellar mass more accurately, since they typically inject more energy (per unit stellar mass formed) into star-forming regions in low-mass galaxies (which dominate at high redshift), than is the case for FBconst. However, the FB$\sigma$ and FBZ models remain inconsistent with the observational measurements at $z>1$, and only the Ref model broadly reproduces these out to $z \sim 2$.

\subsubsection{Specific star formation rates}
\label{sec:ssfr_calibrated}

\begin{figure}
\includegraphics[width=\columnwidth]{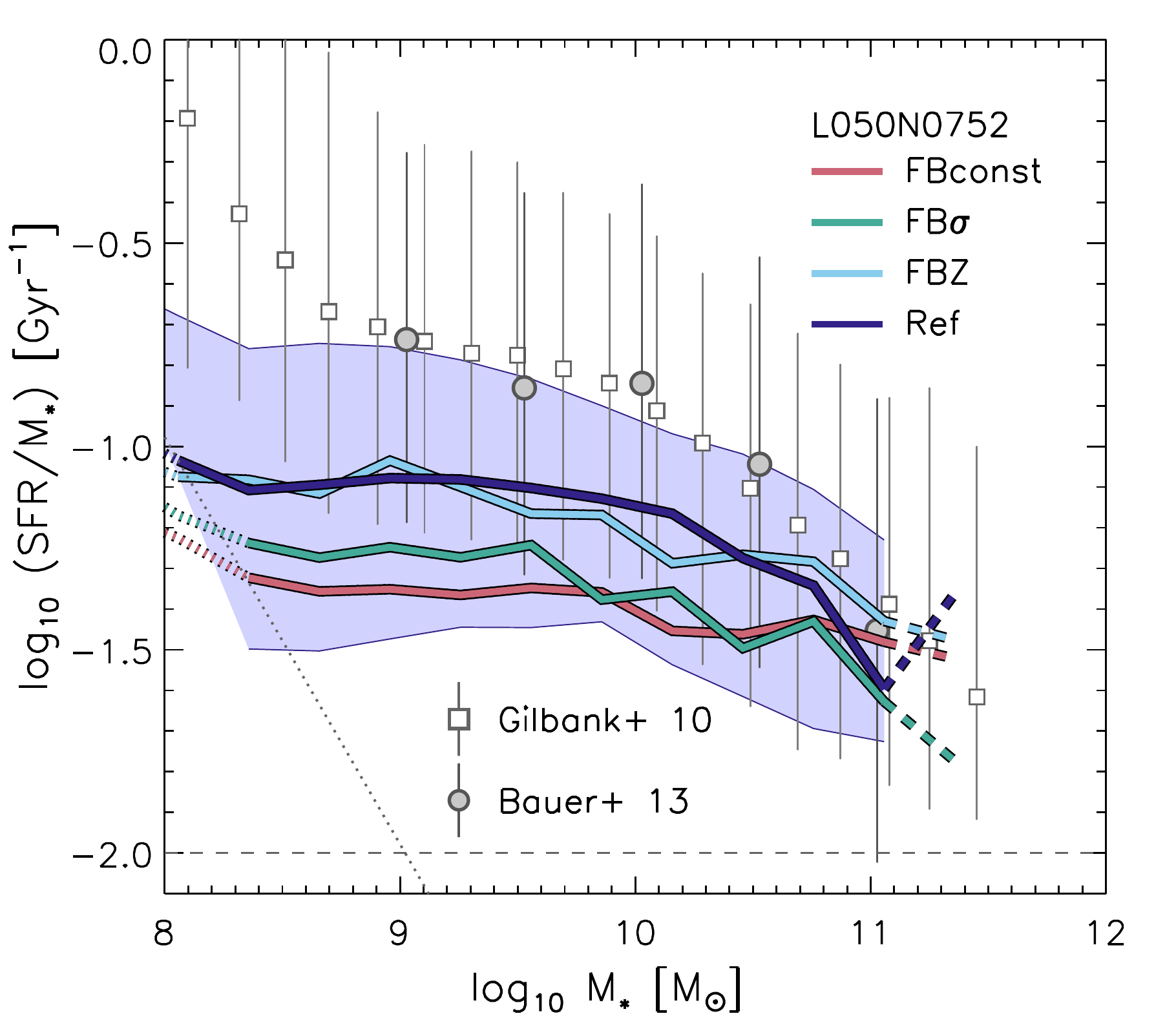}
\caption{The median specific star formation rate, $\dot{M}_\star/M_\star$, of star-forming galaxies as a function of stellar mass at $z=0.1$, of the calibrated EAGLE simulations. The diagonal dotted line corresponds to the SSFR of 10 gas particles at $n_{\rm H}=0.1\,\cmcubed$, below which sampling effects are significant and medians are drawn with a dotted linestyle. The $1\sigma$ scatter about the median of Ref is denoted by the blue shaded region. The dashed horizontal line denotes the SSFR separating star-forming and passive galaxies. Data points with error bars  correspond to the median and $1\sigma$ scatter of the SSFR of star-forming galaxies inferred from SDSS-Stripe 82 data by \citet[][squares]{Gilbank_et_al_10} and GAMA data by \citet[][circles]{Bauer_et_al_13_short}.}
\label{fig:ssfr_calibrated}
\end{figure}

The markedly different evolution of the comoving stellar mass density in the calibrated simulations is indicative of similar differences between the models in terms of the star formation rates of star-forming galaxies at $z=0.1$. The relation between the SSFR ($\dot{M}_\star/M_\star$) and stellar mass at $z=0.1$ is adopted as the diagnostic with which to compare the models; the evolution of the SSFR of Ref is explored by \citet{Furlong_et_al_14_short}. Figure \ref{fig:ssfr_calibrated} shows the median SSFR of star-forming galaxies, as a function of stellar mass, at $z=0.1$ for the four calibrated simulations. As per Figure \ref{fig:mf_calibrated}, the median curve is drawn with a dashed linestyle when sampled by fewer than 10 galaxies per bin. The diagonal dotted line indicates the SSFR corresponding to 10 star-forming gas particles at a gas density of $n_{\rm H} = 0.1\,\cmcubed$; at SSFRs below this limit, sampling limitations are significant and median curves are drawn with a dotted linestyle. In the regime between the limits of sufficient resolution and adequate galaxy sampling, the $1\sigma$ scatter about the median is shown for Ref as a blue shaded region. The horizontal dashed line denotes the threshold SSFR of $10^{-2}~{\rm Gyr}^{-1}$, that was adopted to separate star-forming from passive galaxies. Data points represent SSFR measurements of star-forming galaxies inferred from SDSS-Stripe 82 data \citep[][squares]{Gilbank_et_al_10} and the GAMA survey \citep[][circles]{Bauer_et_al_13_short}. 

S15 demonstrated that, at $M_\star \sim 10^{11}\Msun$, the Ref-L100N1504 simulation exhibits SSFRs very similar to those observed, but at lower masses the simulated SSFRs are systematically lower than observed by up to $\sim 0.3\,{\rm dex}$. This remains the case for Ref-L050N0752. However, as shown by S15, the discrepancy for low mass galaxies is much smaller for the Recal-L025N0752 simulation, indicating that our high-resolution simulations are better able to reproduce the star formation properties of low-mass galaxies. The FBZ model behaves similarly to Ref. The FBconst and FB$\sigma$ models also exhibit SSFRs similar to those observed in massive galaxies, but at $10^9\Msun$ they are $\sim 0.7\,{\rm dex}$ lower than observed. 

In the framework of equilibrium galaxy formation models \citep[e.g][]{Finlator_and_Dave_08,Schaye_et_al_10_short,Dave_Finlator_and_Oppenheimer_12,Mitra_Dave_and_Finlator_14}, the gas inflow rate onto galaxies is balanced by the combined sinks of star formation and ejective feedback, and the specific inflow rate (at fixed redshift) which is a weak function of halo mass \citep{Dekel_et_al_09_short,Fakhouri_Ma_and_Boylan_Kolchin_10,vandeVoort_et_al_11,Correa_et_al_14}. Since the calibrated simulations each broadly reproduce the observed GSMF, galaxies of fixed stellar mass occupy similarly massive haloes in each case: at $z=0.1$, the median halo mass associated with galaxies of stellar mass $M_\star = 10^9\Msun$ in the Ref simulation is offset from that of the FBconst simulation by $< 0.1\,{\rm dex}$. This leaves differences in the mass reaccretion rate and the efficiency of preventive feedback as prime candidates for establishing an offset in the present-day SSFR of low-mass galaxies. 

It is indeed likely that the reaccretion of ejected gas is sensitive to the details of the feedback \citep[e.g.][]{Oppenheimer_et_al_10,Brook_et_al_14} and almost certainly plays a role in shaping the present-day SSFR of galaxies, particularly so at the mass scale corresponding to $M^\star$. We intend to explore this process in detail in a forthcoming study (Crain et al. \textit{in prep}). The efficiency of preventive feedback is, by construction, a distinguishing feature of the four calibrated models, and one that is simple to explore. Examination of the median value of $f_{\rm th}$ associated with star formation events over the gigayear preceding $z=0.1$ for galaxies of $M_\star \sim 10^{9}\Msun$ shows marked differences: for the FBconst, FB$\sigma$, FBZ and Ref models, the values are 1, 1.04, 0.58 and 0.35, respectively. The star formation rate required to produce sufficiently strong outflows from star formation feedback scales inversely with these efficiencies, and thus the Ref model correspondingly exhibits the highest SSFR at $z=0.1$.

\subsubsection{The birth conditions of stars}
\label{sec:birth_conditions_calibrated}

\begin{figure}
\includegraphics[width=\columnwidth]{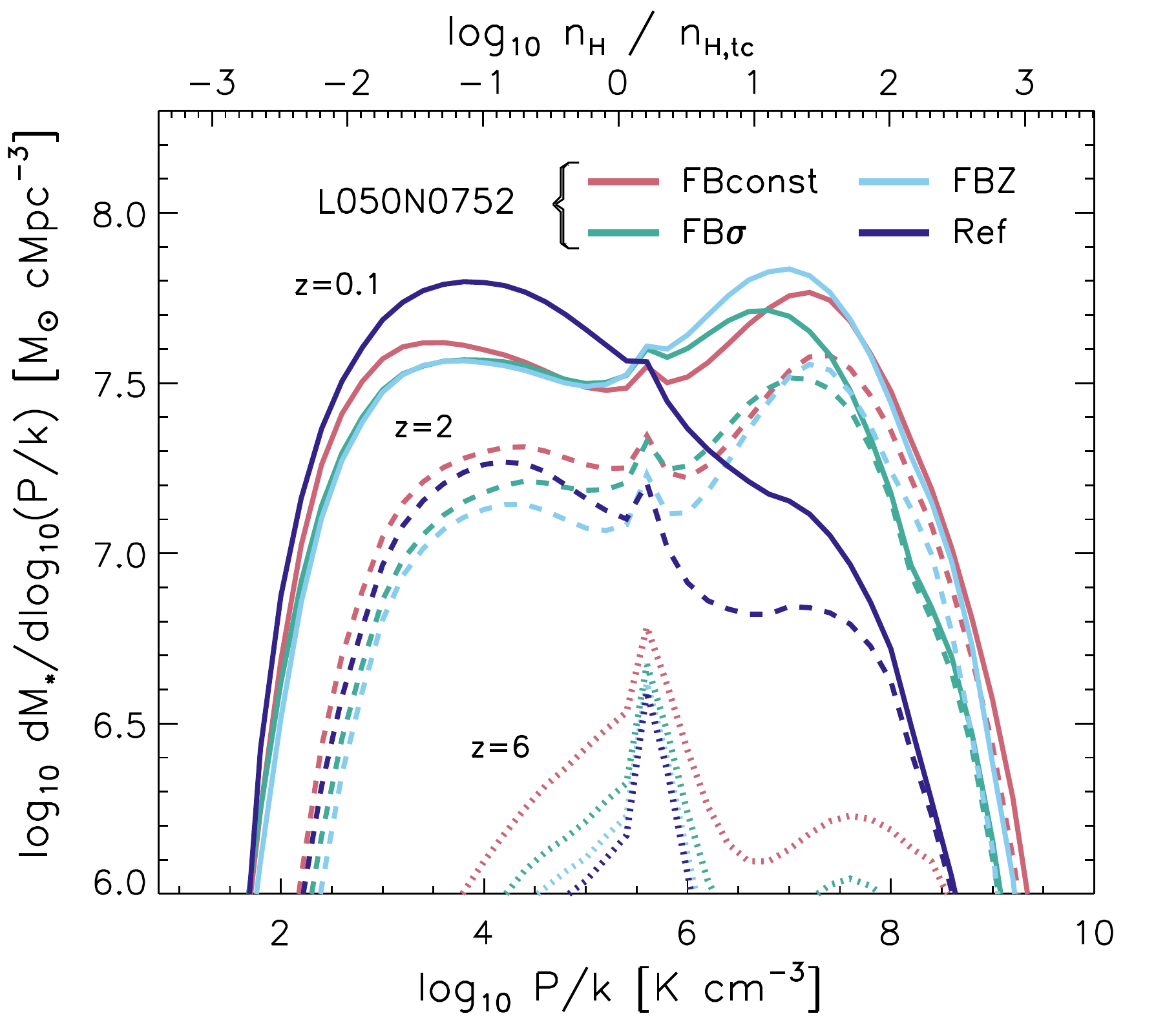}
\caption{Differential distribution of the pressure of fluid elements (i.e. SPH particles) at the instant they were converted to star particles, for all star particles formed prior to $z=6$ (dotted curves), $z=2$ (dashed) and $z=0.1$ (solid). The upper axis shows the ratio of the corresponding density to the critical density for numerically efficient feedback (equation \ref{eq:nhtc}). At high redshift, most stars form from metal-poor gas with a pressure corresponding to the maximum star formation threshold (equation \ref{eq:sfthreshz}), $P/k \sim 10^{5.5}\,{\rm K\,cm}^{-3}$, though in the FBconst model a significant fraction of stars has already formed from gas with $n_{\rm H} \gg n_{{\rm H},t_{\rm c}}$. As the ISM becomes enriched, the threshold drops and stars are able to form at lower ISM densities. However at $z\lesssim 6$ spurious numerical losses in the FBconst, FB$\sigma$ and FBZ simulations result in the rapid build up of a second peak at high-pressure. The development of this peak is suppressed in the Ref simulation owing to the greater efficiency of feedback at higher density (for fixed metallicity), which ensures that the majority of stars are able to yield numerically efficient feedback.}
\label{fig:birth_pressures}
\end{figure}

The properties of simulated galaxies are clearly sensitive to the adopted functional form of $f_{\rm th}$. Galaxy sizes, which encode information related to the state of the gas from which stars were born, are the clearest discriminator of the models explored here, indicating a connection between the stucture of the ISM and the efficacy of feedback. 

When star particles are born, we record the density of their parent gas particle, enabling an examination of the physical conditions of the gas from which all stars in the simulations were born. The EAGLE simulations treat star-forming gas as a single-phase fluid, therefore the SPH density of star-forming particles can be considered as the mass-weighted average of the densities of cold, dense molecular clouds and of the warm, ionised medium with which they maintain a pressure equilibrium. Pressure is therefore a more physically meaningful property of star-forming gas in the simulations, and it is possible to recover the birth pressure of stars from their birth density under the reasonable assumption that their parent gas particle resided on the Jeans-limiting pressure floor at the time of conversion\footnote{Particles within $0.5~{\rm dex}$ of the temperature associated with the pressure floor (see \S\,\ref{sec:ism_and_sf}) are eligible to form stars.}.

Figure \ref{fig:birth_pressures} shows the differential distribution of ISM pressures (normalised by Boltzmann's constant, $k$) at the instant of their formation, of star particles formed in the calibrated simulations prior to $z=6$ (dotted curves), $z=2$ (dashed curves) and $z=0.1$ (solid curves). The upper axis indicates the corresponding ratio of the gas particle density to the critical density for which stochastic thermal heating associated with star formation is efficient ($n_{\rm H,t_c}$, equation \ref{eq:nhtc}). Examination of this ratio affords us a means by which to test for numerical overcooling on an \textit{event-by-event basis}. At high redshift, most stars form from low-metallicity gas, and are hence subject to a high-star formation density threshold (equation \ref{eq:sfthreshz}). The maximum of this threshold is $n_{\rm H} = 10\,\cmcubed$, corresponding to  $P/k \sim 10^{5.5}\,{\rm K\,cm}^{-3}$ for our choice of Jeans limiting equation of state. Many stars form from gas with pressures close to this threshold value.

A significant fraction of stars also form from higher-pressure gas in the FBconst simulation prior to $z=6$. The formation of stars from gas with $n_{\rm H} > n_{{\rm H},t_{\rm c}}$ leads to artificial radiative losses. As discussed in \S\ref{sec:calibrating_fth}, the fact that the first galaxies whose formation can be captured by the simulations are associated with haloes that have not been subject to feedback, means that they exhibit artificially high gas fractions and star formation efficiencies. This initial problem has the potential to set in train a cycle of overcooling: the artificially rapid initial formation of stars over-enriches the ISM with efficient coolants, promoting further cooling losses and enabling dissipation to higher densities. Stars subsequently forming from this gas yield numerically inefficient thermal feedback (because $n_{\rm H} > n_{{\rm H},t_c}$), so the gas fraction and star formation efficiency of the halo remain artificially high. An initial numerical shortcoming therefore has the potential to trigger unrealistic physical losses that themselves promote further numerical losses. This cycle can lead to a strong overestimate of the severity of radiative losses.

\begin{figure}
\includegraphics[width=\columnwidth]{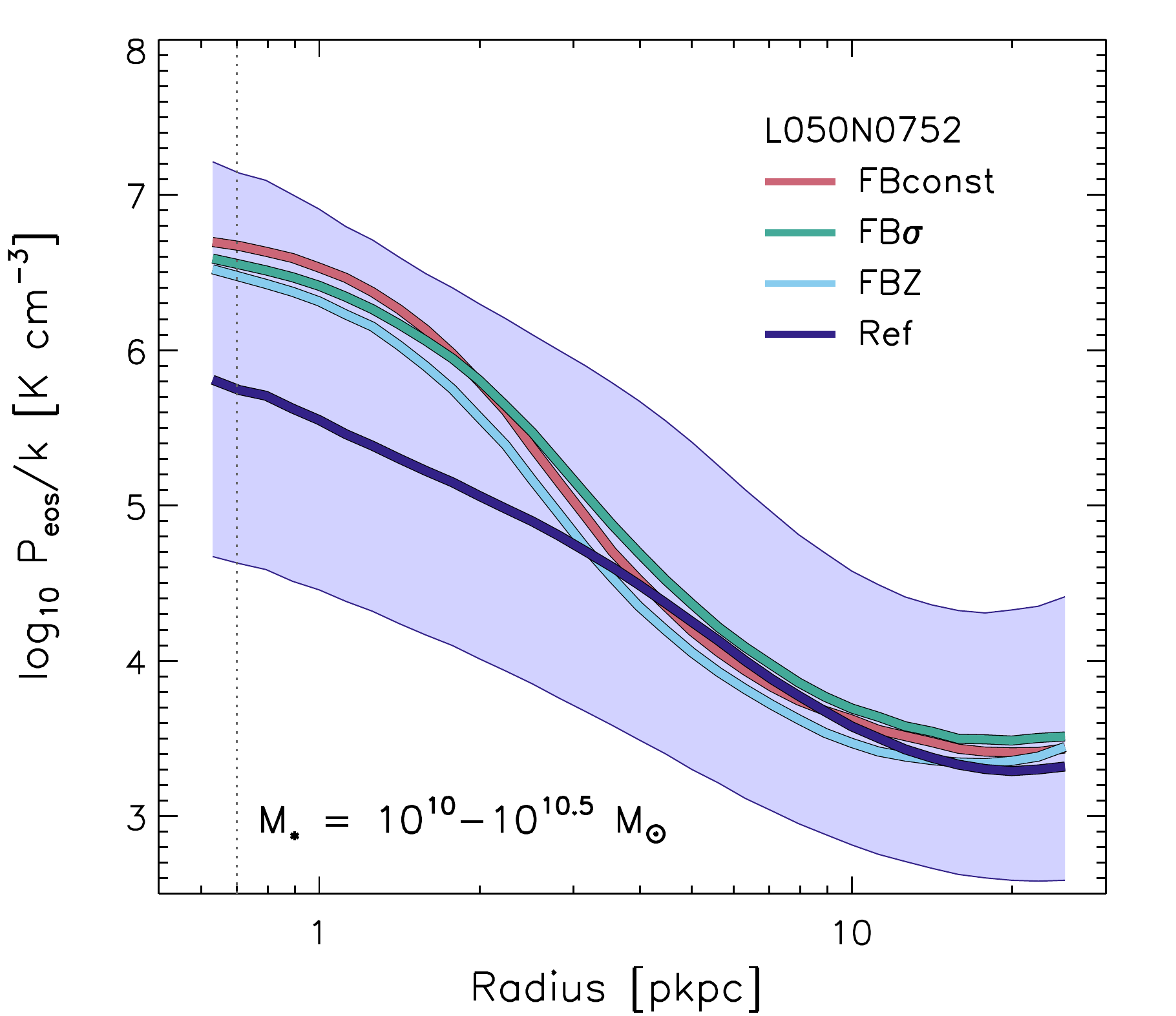}
\caption{The birth pressure of star particles, as a function of galactocentric radius, for galaxies with stellar mass $10 < \log_{10}{M_\star/\Msun} < 10.5$ at $z=0.1$. Solid curves denote the binned median, and the blue shaded region corresponds to the $1\sigma$ scatter about the median of Ref. The dotted vertical line shows the gravitational softening scale. The highest pressures correspond to stars formed in galactic centres, so numerically inefficient feedback in the FBconst, FB$\sigma$ and FBZ models leads to the formation of unrealistically massive central spheroids. Conversely, the suppression of high birth densities in Ref yields effective radii that are consistent with observations. The convergence of the birth pressure profiles for $r \gtrsim 4\pkpc$ indicates that differences in the heating rate of gas (i.e. $f_{\rm th}$) between the models are confined to galactic centres.}
\label{fig:radial_birth_pressures}
\end{figure}

The adoption of $f_{\rm th}^{\rm max}=3$ for stars forming in low-velocity dispersion (FB$\sigma$) and low-metallicity (FBZ, Ref) environments effectively eliminates the initial phase of the problem; at $z=6$, by which time the simulations comprise tens of thousands of star particles, the fraction of stars formed from high-pressure gas is small for the FB$\sigma$, FBZ and Ref simulations. As the ISM becomes enriched with metals, the typical star formation threshold drops and a peak in the distribution of birth densities develops at $P/k \sim 10^{3.5}\,{\rm K\,cm}^{-3}$ ($n_{\rm H} \sim 0.3\,\cmcubed$) in each of the calibrated simulations. This injection of additional energy into nascent galaxies is, however, insufficient to arrest the onset of subsequent numerical losses. At $z \lesssim 6$, the birth pressure distribution of the FB$\sigma$ and FBZ simulations (in addition to that of FBconst) develops a second peak at $P/k \sim 10^{7.5}\,{\rm K\,cm}^{-3}$, corresponding to $n_{\rm H} \sim 250\,\cmcubed$. At such high density, resolution elements heated to $10^{7.5}\K$ cool before they can expand, rendering thermal feedback numerically inefficient. The fraction of stars formed from gas with $n_{\rm H} > n_{{\rm H},t_{\rm c}}$ offers a simple estimate of the severity of this numerical overcooling; for stars formed prior to $z=0.1$ the fractions for the FBconst, FB$\sigma$, FBZ and Ref simulations are 0.55, 0.56, 0.61 and 0.25, respectively.

The suppression of the high-pressure peak in the distribution of birth pressures exhibited by the Ref simulation is achieved by adding a density-dependence to the star formation feedback efficiency function (equation \ref{eq:fth(Z,n)}). It ensures that star formation feedback remains efficient when stars form from relatively dense gas, preventing the build-up of the highest pressures within the ISM. A density dependence of this sort can also be motivated on physical grounds \citep[e.g.][]{Heiles_90,Creasey_Theuns_and_Bower_13,Krause_et_al_13,Nath_and_Shchekinov_13,Roy_et_al_13,Keller_et_al_14}, but our main motivation is to combat the numerical problems described above. 

Figure \ref{fig:radial_birth_pressures} shows the median birth pressure of stars as a function of their galactocentric radius at $z=0.1$, for galaxies with stellar mass in the interval $10^{10} < M_\star/\Msun < 10^{10.5}$. This interval corresponds to the mass scale at which the difference in sizes between Ref and the FBconst, FB$\sigma$ and FBZ models is greatest. The blue shaded region shows the $1\sigma$ scatter about the median of Ref, and the vertical dotted line denotes the gravitational softening scale of $0.7\,{\rm pkpc}$. The correpondence between numerically inefficient feedback and the formation of unrealistically compact galaxies is clear: the highest birth pressures are exhibited by stars residing in the centres of galaxies. The reduction of star formation from gas with $P/k \gtrsim 10^{5.5}\,{\rm K\,cm}^{-3}$ in the Ref simulation therefore preferentially suppresses the formation of compact galactic bulges, and enables the formation of galaxies with effective radii that are consistent with the observed size-mass relation. The median birth pressure profiles converge above a characteristic radius of $r \sim 4\pkpc$, indicating that the difference in heating rates between the simulation (due to differences in the adopted functional form of $f_{\rm th}$) is only significant within the centres of galaxies.

\subsection{Variations of the Reference model}
\label{sec:reference_variations}

The Ref model demonstrates that it is possible to calibrate a model that satisfactorily reproduces the GSMF and the observed sizes of galaxies at $z=0.1$. It is important to quantify the sensitivity of the outcomes of this model to variation of its key subgrid parameters\footnote{The effects of changes to the hydrodynamics and time-stepping schemes will be explored in a forthcoming companion paper (Schaller et al. in prep).}. This is achieved by conducting a series of simulations within which the value of a single parameter is varied from that adopted by Ref, as listed in the lower section of Table \ref{tbl:simulations}. Parameters governing the ISM, star formation and the efficiency of star formation feedback are tested using relatively inexpensive L025N0376 simulations. Those governing the AGN feedback are tested with L050N0752 simulations, since the effects of changing these parameters are most clearly imprinted upon the properties of massive galaxies and their environments. The effects of reasonable changes can vary markedly from parameter to parameter, so we focus here only on properties of the galaxy population that shift significantly from those of the corresponding Ref simulation. Simulations where the variation does have a significant effect are likely to yield a galaxy population that is no longer an accurate representation of the observed Universe.

\subsubsection{ISM equation of state}
\label{sec:vary_gamma}

\begin{figure}
\includegraphics[width=\columnwidth]{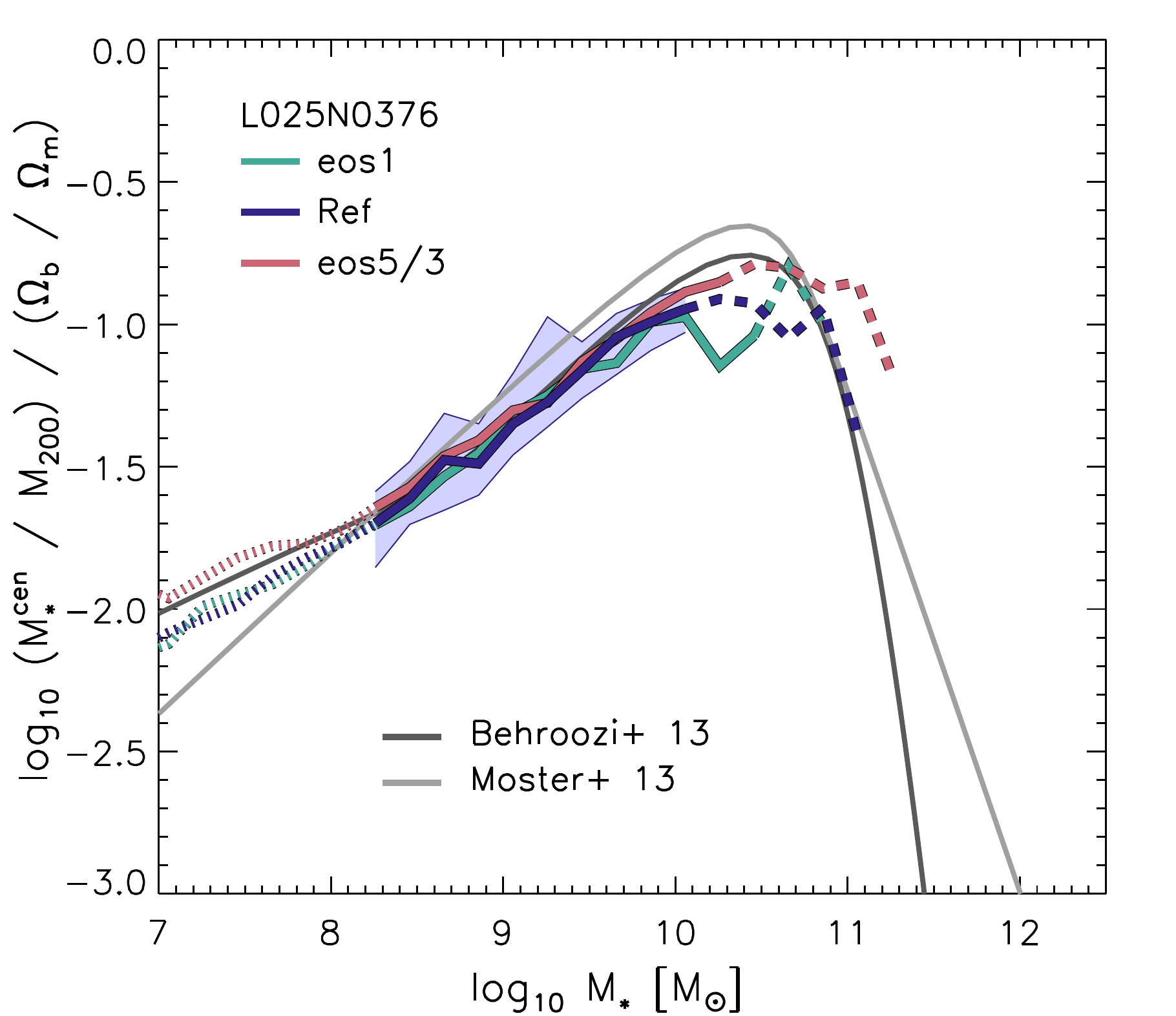}
\caption{The ratio of the stellar mass to halo mass of central galaxies, as a function of stellar mass and normalised by the cosmic baryon fraction, in L0025N0376 simulations adopting $\gamma_{\rm eos}=1$ (eos1, \textit{green}), $4/3$ (Ref, \textit{dark blue}) and $5/3$ (eos5/3, \textit{red}). Curves show the binned median ratios, and are drawn with dotted lines below the mass scale corresponding to 100 baryonic particles, and a dashed line where sampled by fewer than 10 galaxies per bin. The $1\sigma$ scatter about the median of Ref is denoted by the blue shaded region. Dark and light grey lines represent the abundance matching relations of \citet{Behroozi_Wechsler_and_Conroy_13} and \citet{Moster_Naab_and_White_13}, respectively.}
\label{fig:eta_mstar_eos}
\end{figure}

The simulations adopting isothermal ($\gamma_{\rm eos}=1$, ``eos1'') and adiabatic ($\gamma_{\rm eos}=5/3$, ``eos5/3'') equations of state are examined in this section. Steeper slopes, i.e. higher values of $\gamma_{\rm eos}$, yield higher ISM pressures as the volume density is increased, and hence a shorter gas consumption time at fixed density (assuming a star formation law with $n > 1$, which is the case here) and a suppression of the development of high ISM densities. For $4/3<\gamma_{\rm eos}<2$, the Jeans conditions indicate that gas is able to collapse without fragmenting. Steeper slopes inhibit collapse, whilst shallower slopes may lead to artificial fragmentation \citep[e.g.][]{Bate_and_Burkert_97,Schaye_and_Dalla_Vecchia_08}. 

The effects of varying $\gamma_{\rm eos}$ were previously explored in idealised disc galaxy simulations by \citet{Schaye_and_Dalla_Vecchia_08} and in the OWLS cosmological simulations by \citet{Schaye_et_al_10_short} and \citet{Haas_et_al_13b}. These studies concluded that, as long as the index was maintained between values of unity and $5/3$, the primary effect of varying $\gamma_{\rm eos}$ is upon the visual appearance of the disc, with the higher pressures associated with stiffer equations of state yielding smoother gas distributions and larger scale heights. Otherwise, differences due to variations in $\gamma_{\rm eos}$ were found to be minimal, and this was attributed to the fact that the pressure-law implementation of star formation adopted by OWLS (and EAGLE, \S\,\ref{sec:ism_and_sf}) reproduces the observed star formation law, irrespective of $\gamma_{\rm eos}$. 

Figure \ref{fig:eta_mstar_eos} shows the ratio of the stellar mass of central galaxies to the mass of their host halo, normalised by the cosmic average baryon fraction ($\Omega_{\rm b}/\Omega_{\rm m} = 0.157$), and plotted as a function of stellar mass. The dark and light grey lines represent the median stellar mass-to-halo mass ratios inferred from the abundance matching algorithms of \citet{Behroozi_Wechsler_and_Conroy_13} and \citet{Moster_Naab_and_White_13}, respectively. For low-mass galaxies, the conclusion that the properties of galaxies are largely insensitive to $\gamma_{\rm eos}$ is consistent with EAGLE. Changing $\gamma_{\rm eos}$ does, however, have a significant impact on the most massive galaxies since, in contrast to the OWLS reference model, the EAGLE reference model includes treatments of the growth of BHs and the associated AGN feedback. In the regime that the local sound speed is large compared to the relative velocity of the central BH and the surrounding gas ($c_{\rm s} \gg v$), the accretion rate (equation \ref{eq:mdotaccr}) onto the BH is inversely proportional to the cube of the ISM sound speed. For a polytropic equation of state, the sound speed $c_{\rm s}^2 \propto P/\rho \propto \rho^{\gamma_{\rm eos}-1}$, so it is clear that a stiffer equation of state for the ISM will yield a greater sound speed at fixed density, and so reduce the accretion rate onto the BH.

Adopting an isothermal (adiabatic) equation of state leads to galaxies with $M_\star \gtrsim 2 \times 10^{10}\Msun$ exhibiting a median BH mass that is approximately $0.3~{\rm dex}$ larger (smaller) than that of Ref. This difference in BH mass, and the associated difference in the rates of BH accretion and AGN feedback, translates into significant differences in the mass of gas converted to stars by $z=0.1$ in halos of mass $M_{200} \gtrsim 10^{12}\Msun$, as shown in Figure \ref{fig:eta_mstar_eos}. The efficiency of AGN feedback is therefore sensitive to the assumed polytrope $\gamma_{\rm eos}$, and variation of the latter would require a recalibration of the subgrid BH viscosity parameter, $C_{\rm visc}$, to recover the observed GSMF\footnote{In principle, it is possible to construct a model of BH growth that, like the star formation implementation (equation \ref{eq:sflaw}), is independent of $\gamma_{\rm eos}$. This would require that the accretion rate be specified as a function of pressure.}.

\subsubsection{Star formation threshold}
\label{sec:vary_sfthresh}

A simulation adopting a constant density threshold for star formation of $n_{\rm H}^\star = 0.1\,\cmcubed$ (``FixedSfThresh''), as used by default in the OWLS and GIMIC projects, is used to examine the role of the dependence of the star formation density threshold, $n_{\rm H}^\star(Z)$ on metallicity, as implemented in Ref. The use of constant and metallicity-dependent density thresholds for star formation was also investigated using the OWLS simulations by \citet{Schaye_et_al_10_short} and \citet{Haas_et_al_13b}. As in that study, the low-redshift galaxy population exhibits no significant differences between simulations run with a constant density threshold of $n_{\rm H}^\star = 0.1\,\cmcubed$ and the metallicity-dependent threshold described by equation \ref{eq:sfthreshz}, therefore we do not show comparisons of these simulations here. This can be understood by appealing to self-regulation arguments \citep{Schaye_et_al_10_short}.

\subsubsection{Star formation feedback efficiency}
\label{sec:vary_fth}

\begin{figure*}
\includegraphics[width=\textwidth]{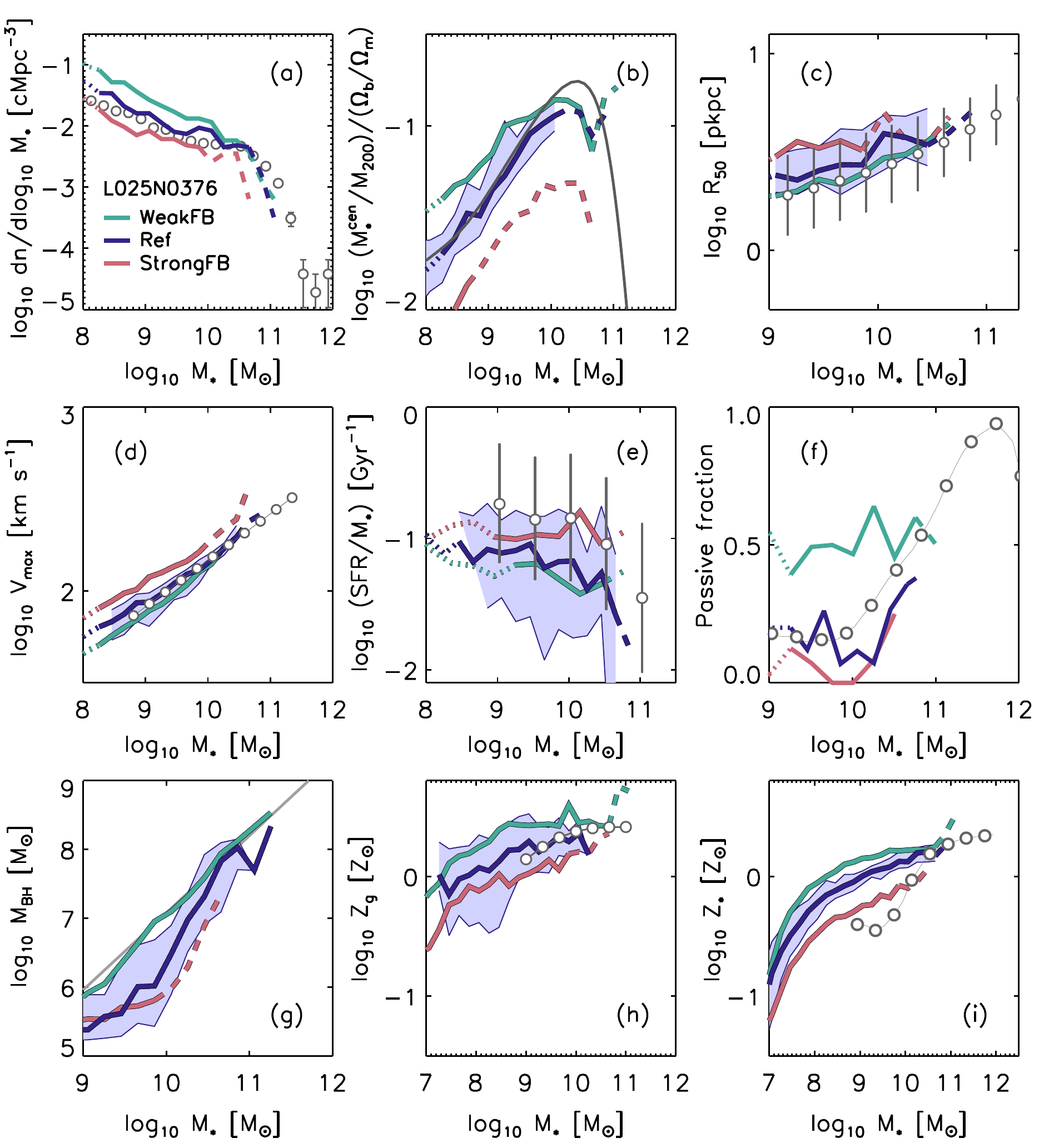}
\caption{Galaxy scaling relations at $z=0.1$ of L0025N0376 simulations adopting star formation feedback decribed by equation \ref{eq:fth(Z,n)} (Ref, \textit{dark blue}), and this function scaled by factors of 0.5 (WeakFB, \textit{green}) and 2  (StrongFB, \textit{red}). With the exception of panels (a) and (f), the curves show binned medians, and the blue shaded region corresponds to the $1\sigma$ scatter about the median of Ref. Curves are drawn with dotted lines below the estimated resolution limit of each diagnostic, and with dashed lines where sampled by fewer than 10 galaxies per bin. The panels show (a) the galaxy stellar mass function; (b) the stellar mass to halo mass ratio of central galaxies; (c) the stellar half-mass radius, (d) the maximum circular velocity, (e) the specific star formation rate of star-forming galaxies; (f) the fraction of galaxies that are passive; (g) the central BH mass; (h) the metallicity of star-forming gas, and (i) the metallicity of stars. The source of the observational data (\textit{dark grey}) for each panel is given in \S\,\ref{sec:vary_fth}.} 
\label{fig:panel9_fth}
\end{figure*}

There is a growing consensus, based on analyses of cosmological simulations, that galaxy formation and evolution is governed primarily by the regulation of the supply of gas to the ISM, rather than the detailed physics of interstellar gas itself \citep[e.g.][]{Rasera_and_Teyssier_06,Bouche_et_al_10_short,Schaye_et_al_10_short,Dutton_van_den_Bosch_and_Dekel_10,Dave_Finlator_and_Oppenheimer_12,Haas_et_al_13a,Haas_et_al_13b,Vogelsberger_et_al_13}. In general, it is feedback that regulates this fuel supply, so it is reasonable to expect that changing the efficiency of star formation feedback will have a significant impact on many characteristics of the galaxy population. This can be gauged by inspection of Figure \ref{fig:panel9_fth}, which shows the effect, on a broad range of galaxy scaling relations, of scaling the $f_{\rm th}$ function adopted by the Ref model (equation \ref{eq:fth(Z,n)}) by factors of $0.5$ (WeakFB) and $2$ (StrongFB). The asymptotic efficiencies of these models are therefore $f_{\rm th}^{\rm min} = (0.15,0.6)$ and $f_{\rm th}^{\rm max} = (1.5,6.0)$, respectively. The median efficiency associated with all star particles formed prior to $z=0.1$ in the WeakFB, Ref and StrongFB L0025N0376 simulations are 0.38, 0.63 and 1.22, respectively. 

As in the previous plots, binned medians are shown (WeakFB: green curve, Ref: dark blue curve, StrongFB: red curve), drawn with a dotted linestyle below the estimated resolution limit, specific to each diagnostic, and a dashed linestyle where there are fewer than 10 galaxies per bin. Where appropriate, the $1\sigma$ scatter about the median of Ref is shown as a blue shaded region. The panels are labelled (a) to (i) in row-by-row and top-to-bottom order, showing (source of observational data in parentheses): (a) the GSMF \citep{Baldry_et_al_12_short}; (b) the stellar mass to halo mass ratio of central galaxies \citep{Behroozi_Wechsler_and_Conroy_13}; (c) the stellar half-mass radius \citep{Shen_et_al_03}, (d) the maximum circular velocity, which we adopt as a proxy for the Tully-Fisher relation \citep{Avila_Reese_et_al_08}, (e) the specific star formation rate of star-forming galaxies \citep{Bauer_et_al_13_short}; (f) the fraction of galaxies that are passive \citep{Moustakas_et_al_13_short}; (g) the central BH mass \citep{McConnell_and_Ma_13}; (h) the metallicity of star-forming gas \citep{Zahid_et_al_14}, and (i) the metallicity of stars \citep{Gallazzi_et_al_05}. For more details concerning the observations and their comparison with the simulations, see S15.

Inspection of the panels highlights that relatively small changes (i.e. factor 2) to the value of $f_{\rm th}$ adopted by Ref has a dramatic impact upon the properties of the $z=0.1$ galaxy population. Panel (a) shows that a lower (higher) star formation feedback efficiency corresponds to a greater (lesser) abundance of galaxies with masses below $M^\star$, the scale corresponding to the ``knee'' of the \citet{Schechter_76} function. The cause of this shift of number densities is clear from inspection of the stellar mass to halo mass relation, (b). The form of the median relation is similar in all three cases, but with a significant normalisation offset: in the WeakFB (StrongFB) model, galaxies acquire a lower (higher) characteristic value of $f_{\rm th}$ throughout their growth. In the framework of equilibrium models, this requires a higher (lower) SFR to produce sufficiently strong outflows to balance the inflow of gas and, by $z=0.1$, leads to galaxies of a fixed stellar mass becoming associated with less (more) massive dark matter haloes, with respect to Ref. Since the dark matter halo mass function is particularly steep \citep[e.g.][]{Jenkins_et_al_01}, even a small change to the relation between stellar mass and halo mass impacts significantly upon the number density of galaxies at a fixed stellar mass. Even on more massive scales, the adoption of very efficient star formation feedback dramatically reduces the abundance of galaxies, although on these scales the adoption of inefficient star formation feedback does not result in a commensurate increase in the abundance of galaxies. As explored below, in the absence of efficient star formation feedback, BHs simply grow more massive (at fixed halo mass) in order to liberate the energy required to achieve self-regulation.

The shift of the typical halo mass associated with galaxies of fixed stellar mass, as the feedback efficiency is varied, impacts significantly upon galaxy scaling relations, as we explore below. In general, the following panels highlight the importance of populating haloes with galaxies of the correct stellar mass, whether ``by-hand'' in empirical models such as the halo occupation distribution or abundance matching methodologies, or as a result of the implemented baryon physics in semi-analytic and hydrodynamic simulations.

The connection between feedback and the sizes of galaxies was highlighted in Sections \ref{sec:sizes_calibrated} and \ref{sec:birth_conditions_calibrated}. Consistent with that discussion, the impact of the star formation feedback efficiency upon the sizes of galaxies is evident in panel (c). As per Figure \ref{fig:sizes_calibrated}, only galaxies for which the best-fitting S\'{e}rsic profile has index $n_{\rm s}<2.5$ are considered. At fixed stellar mass, the WeakFB (StrongFB) model yields smaller (larger) galaxies, with respect to Ref. The adoption of WeakFB leads to significant overcooling (in a physical sense), and results in artificially compact galaxies. In the regime of less significant overcooling, the efficiency of feedback still impacts upon galaxy sizes: feedback preferentially ejects the lowest angular momentum gas in galaxies \citep[e.g.][]{Brook_et_al_11_short,Brook_et_al_12}, so the heating (and concomitant ejection) of more gas per unit stellar mass formed in the StrongFB model with respect to Ref increases the median angular momentum of the ISM gas that remains to form stars, leading to the formation of more extended galaxies.

Panel (d) shows the maximum circular velocity of haloes, $V_{\rm max} = \max{(\sqrt{GM(<r)/r})}$, as a function of the stellar mass of their central galaxies. This is a close relative of the \citet{Tully_and_Fisher_77} relation exhibited by disc galaxies, so once again only those galaxies with $n_{\rm s} < 2.5$ are plotted. Changing the efficiency of star formation feedback impacts upon both the zero-point and the slope of the relation. The zero-point is affected since, as in (b), for a broad range of halo masses, the primary effect of changing the efficiency is to change the stellar mass that is associated with a given halo mass \citep[see also][]{McCarthy_et_al_12}. To first order $V_{\rm max}$ is a reasonable proxy for halo mass, so changing the star formation feedback efficiency essentially shifts the stellar mass associated with a fixed $V_{\rm max}$, i.e. translates the median curves left-to-right. However, the slope also changes because star formation feedback impacts most significantly upon low-mass galaxies, as is also clear from inspection of (b). Moreover, in high-mass galaxies weak feedback can result in an increase of $V_{\rm max}$ due to the formation of compact bulges.

The SSFR as a function of stellar mass is shown in panel (e), and highlights that the adoption of weaker (stronger) star formation feedback leads to lower (higher) SSFR at fixed stellar mass. At first glance this may appear to contradict the interpretation of the relations presented in Figure \ref{fig:ssfr_calibrated} (see \S\,\ref{sec:ssfr_calibrated}), where we concluded that the weaker feedback at low redshift in Ref (with respect to FBconst, for example) enabled low-mass galaxies to achieve a higher SSFR. There is, however, an important distinction with respect to the calibrated simulations: galaxies of a fixed stellar mass in the WeakFB and StrongFB models are not associated with haloes of a similar mass to their counterparts in Ref. The association of galaxies of fixed stellar mass with less (more) massive haloes in the WeakFB (StrongFB) model leads to them experiencing lower (higher) infall rates (both from the cosmological accretion of intergalactic gas, and the reaccretion of ejected material). Increasing the efficiency of feedback enables galaxies to balance a fixed net inflow rate at a lower SFR, but this is insufficient to compensate for the increased inflow rate that stems from being associated with a more massive halo. A related symptom of the different SSFRs seen when varying the feedback efficiency, is a significant shift of the fraction of galaxies that are passive (SSFR $< 10^{-2}~{\rm Gyr}^{-1}$) at $z=0.1$, as shown in panel (f). The impact of changing the star formation feedback efficiency is not a simple shift in the normalisation of passive fraction as a function of stellar mass, since approximately half of all galaxies in the stellar mass range examined are classified as passive in WeakFB. In contrast, very few galaxies with stellar mass $M_\star \sim 10^{10}\Msun$ are passive in StrongFB, whilst at higher masses the StrongFB model converges on the relation realised by Ref. The adoption of weak star formation feedback results in a dramatically greater passive fraction for two reasons. Firstly, galaxies consume a greater fraction of their low entropy gas for star formation, and do so at early times, reducing the reservoir of cold gas available for star formation at the present epoch. The second, and more important effect, is that changing the efficiency of star formation feedback also impacts upon the relationship between the masses of galaxies and their central BHs, as shown in panel (g). 

The significantly higher (lower) normalisation of the $M_{\rm BH}-M_\star$ relation in WeakFB (StrongFB) with respect to Ref may appear, at first glance, to be counter-intuitive. Using OWLS simulations featuring the \citet{Booth_and_Schaye_09} implementation of AGN feedback, \citet{Booth_and_Schaye_10} concluded that the mass of BHs is established primarily by the mass of their host dark matter halo, with a secondary dependence upon the concentration of the halo. Na\"ively, one might therefore expect that galaxies of fixed stellar mass would host less (more) massive BHs, with respect to Ref, in WeakFB (StrongFB), since they reside in lower- (higher-) mass haloes. However, this reasoning neglects another consequence of changing the star formation feedback efficiency: in order to achieve self-regulation, BHs in the WeakFB (StrongFB) case must compensate for the lower (higher) star formation feedback efficiency by liberating more (less) AGN feedback energy. Since we adopt a fixed feedback efficiency for AGN, $\epsilon_{\rm f}=0.15$, BHs can only adjust their energy output by growing at different rates and hence yielding $M_{\rm BH}-M_\star$ relations that are offset from that of Ref at $z=0.1$. Note that in the WeakFB case, no upturn is seen in the $M_{\rm BH}-M_\star$ relation at intermediate mass scales, since the lack of stellar feedback enables BHs hosted by low-mass galaxies to reach the mass at which their growth becomes regulated, or even quenched, by AGN feedback. 

Finally, panels (h) and (i) show the metallicities of the ISM (specifically, star-forming particles) and stars, respectively, as a function of stellar mass. At the low- and intermediate-mass scales, the simple shift in the normalisation of the relations as a function of the adopted stellar feedback efficiency indicates that the outflows driven by this mechanism eject metal-rich gas from galaxies, preventing galaxies from acquiring the metallicity expected in the simple ``closed box'' chemical evolution scenario. The convergence of the median relations for the three models at high metallicity indicates that the ejection of metal-rich gas in massive galaxies is less efficient than at low masses, and/or is not driven primarily by star formation feedback \citep[see also e.g.][]{de_Rossi_Tissera_and_Scannapieco_07,Dave_Finlator_and_Oppenheimer_11,Obreja_et_al_14, Creasey_Theuns_and_Bower_15}. The observation of a similar anti-correlation between baryonic mass and metal-loss in local galaxies is widely perceived as convincing evidence for the ubiquity of outflows, and of their efficiency as a mechanism to transport metals away from the ISM \citep[e.g.][]{Tremonti_et_al_04_short,Zahid_et_al_14}.

In summary, the efficiency of feedback associated with star formation has a strong effect on a broad range of galaxy scaling relations. This is primarily, but not exclusively, because factor of two changes to the efficiency significantly alter the relationship between stellar mass and halo mass. The changes seen in the properties of the $z=0.1$ EAGLE galaxy population as a result of such changes are, for several key scaling relations, at a level that is significantly greater than the observational uncertainty. 

\subsubsection{Subgrid accretion disc viscosity}
\label{sec:vary_cvisc}

\begin{figure*}
\includegraphics[width=\textwidth]{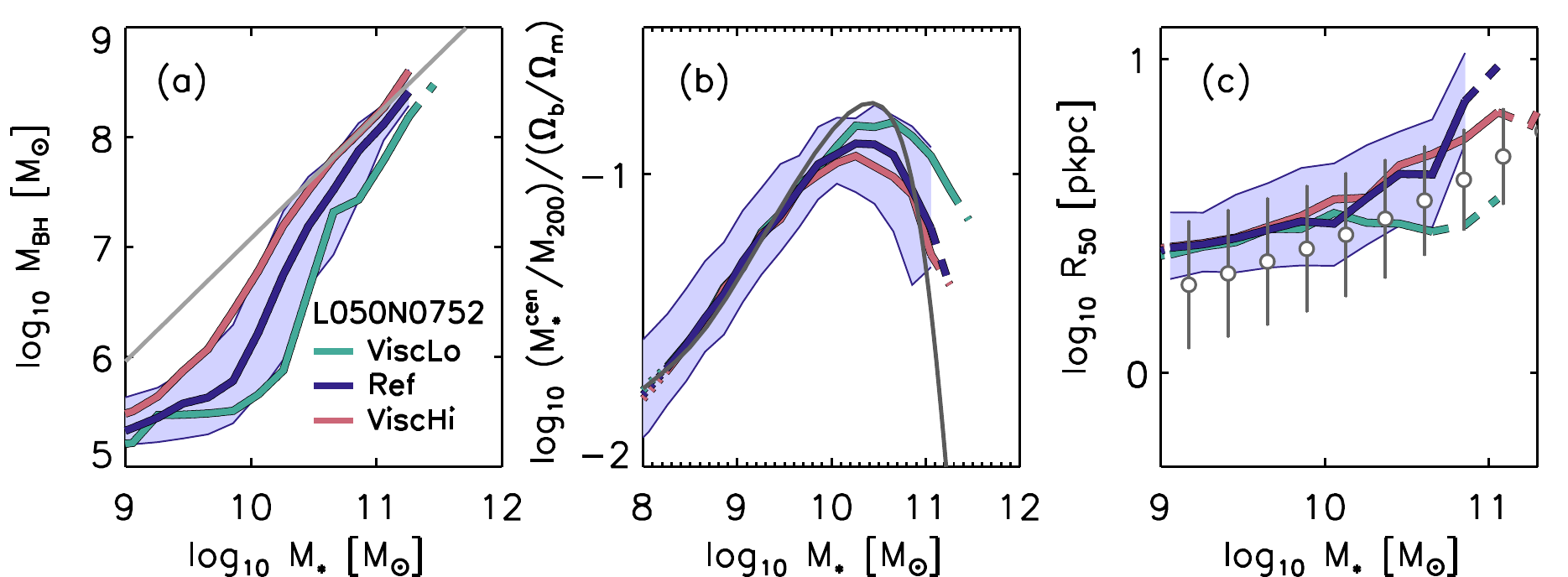}
\caption{Galaxy scaling relations at $z=0.1$ in L0050N0752 simulations adopting $C_{\rm visc}=2\pi \times 10^2$ (ViscLo, \textit{green}), $C_{\rm visc}=2\pi \times 10^{0}$ (Ref, \textit{dark blue}), and $C_{\rm visc}=2\pi \times 10^{-2}$ (ViscHi, \textit{red}). Curves show binned medians, and the blue shaded region corresponds to the $1\sigma$ scatter about the median of Ref. Curves are drawn with dotted lines below the estimated resolution limit of each diagnostic, and dotted lines where sampled by fewer than 10 galaxies per bin. The panels show (a) the $M_{\rm BH}-M_\star$ relation, (b) the stellar mass to halo mass ratio of central galaxies; (c) the effective radius of disc galaxies as a function of stellar mass. Details of the observational measurements (\textit{dark grey}) are provided in \S\,\ref{sec:vary_fth}.}
\label{fig:panel3_agn_visc}
\end{figure*}

The parameter $C_{\rm visc}$ is related to the inverse of the viscosity of a notional subgrid accretion disc, and has two effects. Firstly, it governs the angular momentum scale at which the accretion switches from the relatively inefficient viscosity-limited regime to the Bondi-limited regime (with both cases being subject to the Eddington limit). Secondly, it governs the rate at which gas transits through the accretion disc when the viscosity-limited regime applies. A higher subgrid viscosity, which corresponds to a \textit{lower} value of the viscosity parameter $C_{\rm visc}$, therefore leads to an earlier onset of the dominance of AGN feedback, and a greater energy injection rate when in the viscosity-limited regime. Figure \ref{fig:panel3_agn_visc} shows the $z=0.1$ scaling relations that are most significantly affected as $C_{\rm visc}$ is varied by factors of $10^{2}$ (ViscLo) and $10^{-2}$ (ViscHi), respectively, from the value adopted by Ref. These are (a) the $M_{\rm BH}-M_\star$ relation, (b) the stellar mass to halo mass ratio of central galaxies, and (c) the effective radius as a function of stellar mass.

The $M_{\rm BH}-M_\star$ relation exhibits three distinct regimes: an initial slow and inefficient growth from the seed mass ($10^5\hMsun$), followed by a period of rapid growth of the BH towards the observed high-mass scaling relation and, finally, steadier growth along the scaling relation. The slow initial growth stems from several physical causes. Growth by mergers with seed mass BHs is inefficient, simply because the integrated mass of seeds encountered by any given BH is small \citep[see Figure 4 of ][]{Booth_and_Schaye_09}. Growth by gas accretion is also initially inefficient, firstly because the Bondi rate scales strongly with BH mass ($\propto M_{\rm BH}^2$), and secondly because the rotational support of the ISM in low-mass galaxies is sufficient to maintain the angular momentum of gas close to the BH above the accretion threshold introduced by \citet{Rosas_Guevara_et_al_14_short}. The significance of the latter is clear from inspection of panel (a), which shows that the characteristic mass at which BHs begin to grow efficiently is sensitive to $C_{\rm visc}$. This scale is $M_\star \sim 10^{10}\Msun$ in Ref, whilst in the ViscLo and ViscHi models it is shifted to $M_\star \sim 10^{10.3}\Msun$ and $M_\star \sim 10^{9}\Msun$, respectively.

Once gas accretion is efficient, BHs grow rapidly, because the feedback liberated by their growth is initially unable to regulate the accretion rate. Once sufficiently massive, however, BHs become able to regulate, or even quench, their own growth by gas accretion. In this regime, the gas that does accrete onto BHs has relatively low specific angular momentum \citep{Rosas_Guevara_et_al_14_short}. Therefore the stellar mass at which BHs arrive on the scaling relation is less sensitive to $C_{\rm visc}$ than the stellar mass at which BHs begin to accrete efficiently. The slope of the $M_{\rm BH}-M_\star$ relation in the regime of efficient growth is hence sensitive to $C_{\rm visc}$, being significantly steeper (shallower) than Ref for ViscLo (ViscHi). However, the three simulations do not converge to the same relation, indicating that accretion onto the BH remains (partially) viscosity-limited in even the most massive BHs. In the most massive galaxies, poorly sampled by the $L=50\cMpc$ volumes used here, the growth of BHs is likely dominated by mergers.

The role of the subgrid viscosity as a means to calibrate the simulations is clear from panel (b). By shifting the stellar mass scale at which BHs begin to self-regulate, the assumed viscosity effectively controls the halo mass scale at which AGN feedback becomes significant. The amplitude of the baryon conversion efficiency, and the halo mass scale at which the peak occurs, are therefore sensitive to the viscosity; appealing to lower (higher) viscosity increases (decreases) both the amplitude and halo mass scale of the peak stellar mass to halo mass ratio \citep[see also][]{Rosas_Guevara_et_al_14_short}. 

In the previous section, we concluded that the size of low-to-intermediate mass galaxies is sensitive to the efficiency of feedback associated with star formation (i.e. $f_{\rm th}$). Star formation in more massive galaxies is regulated primarily by AGN feedback, so their sizes are more sensitive to the parameters governing AGN feedback, as shown in panel (c). Delaying the onset of efficient AGN feedback with a low subgrid viscosity results in the delivery of gas to the ISM being countered by star formation feedback alone. A greater fraction of the stars in massive galaxies form from particularly dense, metal-rich gas that, as discussed in \S\,\ref{sec:birth_conditions_calibrated}, is not conducive to the realisation of numerically efficient feedback. Some degree of overcooling is therefore to be expected, a symptom of which is the formation of a massive, compact bulge component that reduces the effective radius of the galaxy. As for adjustments to $f_{\rm th}$, the change in the typical halo mass associated with galaxies of a fixed stellar mass also affects the sizes of galaxies, but here the effect is weaker, and limited to massive galaxies.

\subsubsection{AGN heating temperature}
\label{sec:vary_agndt}

\begin{figure*}
\includegraphics[width=\textwidth]{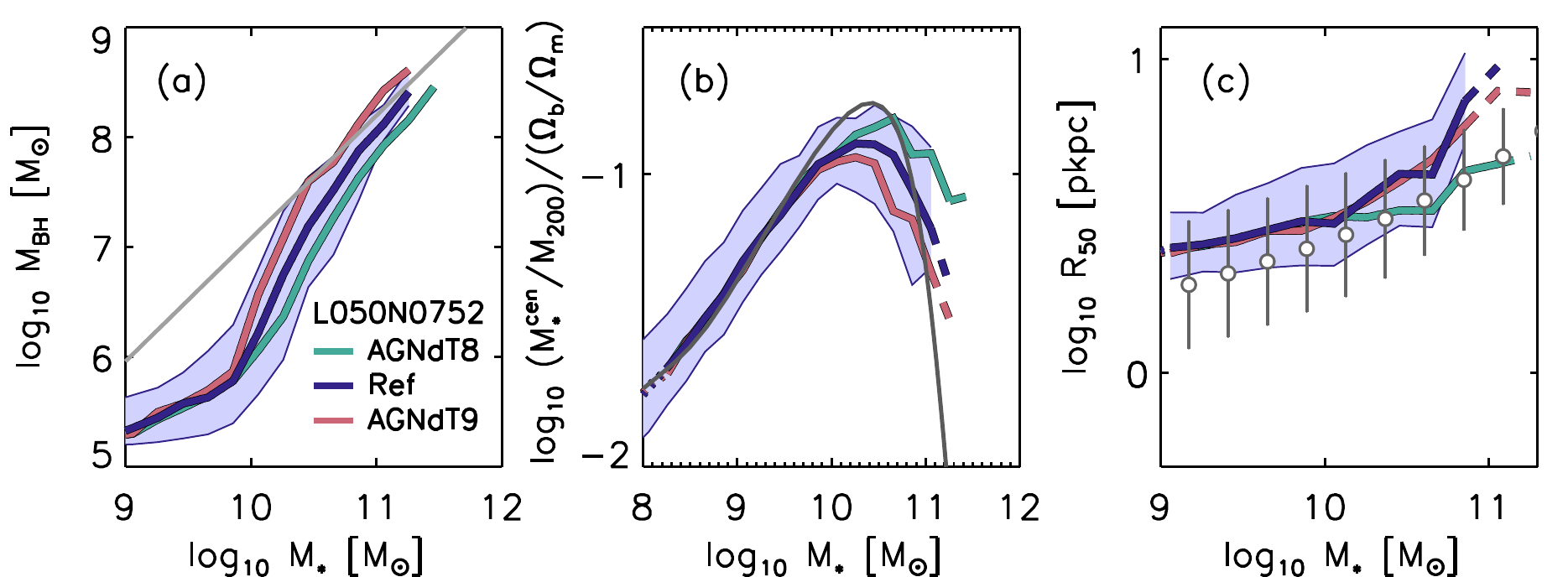}
\caption{As per Figure \ref{fig:panel3_agn_visc}, but varying the AGN heating temperature, $\Delta T_{\rm AGN}$, instead of the subgrid viscosity. The curves correspond to $\Delta T_{\rm AGN}=10^{8}\K$ (AGNdT8, \textit{green}), $\Delta T_{\rm AGN}=10^{8.5}\K$ (Ref, \textit{dark blue}), and $\Delta T_{\rm AGN}=10^{9}\K$ (AGNdT9, \textit{red}).}
\label{fig:panel3_agn_agndt}
\end{figure*}

The crucial role of the AGN heating temperature in EAGLE was explored in part by S15, who presented results from a L0050N0752 simulation adopting $\Delta T_{\rm AGN}=10^9\K$. They concluded that the higher heating temperature, which yields more energetic but less frequent AGN feedback episodes, was necessary to reproduce the gas fractions and X-ray luminosities of galaxy groups. Recently, \citet{Le_Brun_et_al_14} used the cosmo-OWLS simulation suite to conduct a systematic examination of the properties of galaxy groups in response to variation of $\Delta T_{\rm AGN}$. They too concluded that a higher heating temperature yields more efficient AGN feedback.

In Figure \ref{fig:panel3_agn_agndt}, we again show the $M_{\rm BH}-M_\star$ relation, baryon conversion efficiency and the size of galaxies at $z=0.1$, this time for simulations adopting $\Delta T_{\rm AGN}=10^{8}\K$  (``AGNdT8'') and $\Delta T_{\rm AGN}=10^{9}\K$  (``AGNdT9''), besides Ref which adopts $\Delta T_{\rm AGN}=10^{8.5}\K$. The trends appear qualitatively similar to those resulting from changes to the subgrid viscosity, but there are important differences. Variation of the heating temperature does not affect the galaxy stellar mass scale at which BHs begin to grow towards the observed $M_{\rm BH}-M_\star$ relation. Therefore, the halo mass scale at which the baryon conversion efficiency peaks is also independent of the heating temperature. 

However, once a central BH is sufficiently massive to impact upon its environment, the macroscopic efficiency (which should not be confused with the subgrid efficiency, $\epsilon_{\rm f}$) of the associated AGN feedback is sensitive to the adopted heating temperature. The less frequent but more energetic heating events associated with a higher heating temperature are more effective at regulating star formation in massive galaxies, and as a result the peak baryon conversion efficiency is higher (lower) in the AGNdT8 (AGNdT9) model, with repect to Ref. Since the mass of BHs in the simulation is established to first order by halo mass \citep{Booth_and_Schaye_10}, the shift in halo mass associated with galaxies of stellar mass $M_\star \gtrsim 10^{10}\Msun$ establishes the offset in the $M_{\rm BH}-M_\star$ relation above this mass scale.

The behaviour of the size-mass relation in response to variation of the AGN heating temperature, panel (c), is consistent with that resulting from variation of the star formation feedback efficiency (Figure \ref{fig:panel9_fth}) and the subgrid viscosity  (Figure \ref{fig:panel3_agn_visc}). The reduced efficacy of AGN feedback when a lower heating temperature is adopted leads to the formation of more compact galaxies, since a greater fraction of stars are able to form from gas whose density is higher than the critical density for numerically efficient feedback, $n_{{\rm H},t_{\rm c}}$ (equation \ref{eq:nhtc}).

\section{Summary and discussion}
\label{sec:summary_and_discussion}

This study presents results derived from simulations from the ``Evolution and Assembly of GaLaxies and their Environment'' (EAGLE) project. The simulations adopt cosmological parameters that are guided by recent results from the \textit{Planck} satellite, were realised with a force resolution of $0.7\pkpc$ for $z < 3$ (better at higher redshift), and adopt a baryonic particle mass of $1.8 \times 10^6\Msun$. They were conducted with a version of the \gadget\ software that incorporates modified implementations of SPH, time stepping and subgrid models governing cooling, star formation, stellar evolution, feedback from star formation, gas accretion onto and mergers of BHs, and AGN feedback. These simulations complement the three introduced in a companion paper by S15, which also included simulations with higher resolution than the intermediate-resolution simulations presented here. The EAGLE suite of simulations also includes very-high resolution `zoom' simulations of cosmological environments similar to the Local Group \citep{Sawala_et_al_14a_short,Sawala_et_al_14b_short}.

S15 discussed the implications of the crucial role played by subgrid routines in cosmological simulations, concluding that cosmological simulations are at present unable to estimate the efficiency of feedback processes from first principles, and thus cannot predict stellar and BH masses. The optimal recourse is therefore to calibrate the parameters of subgrid routines to ensure that simulations reproduce well-characterised observables, for example the stellar mass function of galaxies. The introduction of several new simulations here enables a clear illustration of the motivation for the parametrisation adopted by the EAGLE reference model. The additional simulations also enable an exploration of the sensitivity of the outcomes of the reference model to the variation of its key subgrid parameters. 

The EAGLE simulations each adopt the stochastic thermal feedback scheme of \citet{Dalla_Vecchia_and_Schaye_12}, whereby fluid elements are stochastically heated to a pre-specified heating temperature, and the appropriate time-averaged energy injection rate is achieved by adjusting the heating probability. This scheme mitigates a common problem in cosmological simulations, namely that the energy injected as feedback at each timestep is usually radiated away before the gas is able to expand. Moreover, this scheme enables the quantity of energy per feedback event to be specified, irrespective of the time-averaged energy injection rate. In the case of feedback associated with star formation, SPH neighbours of newly-formed star particles are heated to $\Delta T_{\rm SF}=10^{7.5}\K$, and the probability of heating is determined by the fraction of the energy budget that is available for feedback, $f_{\rm th}$. We adopt the convention that $f_{\rm th}=1$ corresponds to an expectation value of the injected energy of $1.736\times 10^{49}\,\erg\,\Msun^{-1}$ (or $8.73\times 10^{15}\,\erg\,\g^{-1}$), the energy expected from type II supernovae associated with a Chabrier IMF, assuming each SN liberates $10^{51}\erg$. For this energy budget and heating temperature, $\sim1.3$ fluid elements are expected to be heated by each newly-formed star particle (equation \ref{eq:Nav}).

Four of the simulations presented here, referred to as the `calibrated simulations' (\S\ref{sec:calibrated_simulations}), adopt different functional forms for $f_{\rm th}$. The efficiency of the subgrid star formation feedback that each simulation assumes, and the fashion in which this depends upon the properties of the local environment, is different in each case. The calibrated simulations are realised within a cubic comoving cosmological volume of $L=50\cMpc$ on a side, and each reproduces the observed $z=0.1$ GSMF to within $\lesssim 0.3~{\rm dex}$ over the range $10^{8.2} \lesssim M_\star \lesssim 10^{11.2} \Msun$ (Figure \ref{fig:mf_calibrated}). This precision is similar to that typically attained by semi-analytic galaxy formation models, and is unprecedented for hydrodynamical simulations. The GSMF realised in each case is not presented as a prediction, because the subgrid feedback parameters were calibrated using the observations. Nevertheless, the reproduction of the observed GSMF represents a significant advance. The computational expense of calibrating large hydrodynamical simulations precludes an exhaustive exploration of the available parameter space, so success was not guaranteed, and indeed there was no guarantee that the adopted subgrid models could reproduce the observations for \textit{any} values of their parameters. More practically, the availability of hydrodynamical simulations that reproduce the demographics of the local galaxy population enables a diverse range of problems to be addressed, many of which have previously been limited to investigation with semi-analytic models.

This development adds to recent progress towards the reproduction of the observed $z=0.1$ GSMF (or, in the case of `zoom' simulations, the $M_{\star} - M_{\rm halo}$ relation) in hydrodynamical simulations \citep[e.g.][]{Okamoto_et_al_10,Oppenheimer_et_al_10,Munshi_et_al_13_short,Puchwein_and_Springel_13,Vogelsberger_et_al_13,Hopkins_et_al_14,Vogelsberger_et_al_14_short}. Success in this regard is typically interpreted as an indication that a particular feedback implementation is reasonable. However, the reproduction here of the observed $z=0.1$ GSMF, with a number of models that yield unrealistically compact galaxies, highlights that the simultaneous comparison of simulated galaxies with complementary diagnostics, such as the observed sizes of galaxies, is necessary to ensure the accurate reproduction of a broad range of galaxy scaling relations\footnote{\citet{Hopkins_et_al_14} examined the size of galaxies in the FIRE `zoom' simulations, concluding they are consistent with observations, whilst the Illustris team recently posted a preprint examining their size-mass relation, concluding that their $\sim L\star$ galaxies are larger than observed by a factor of $\sim 2$ \citep{Snyder_et_al_15_short}.}. 

For example, the FBconst model, which adopts $f_{\rm th}=1$ irrespective of local conditions, yields a cosmic star formation history that is weighted too strongly to early cosmic epochs, resulting in the formation of galaxies that are too compact and too quiescent at the present epoch. The dominant cause of the shortcomings of the FBconst model (and also the FB$\sigma$ and FBZ models) is spurious, numerical radiative losses that artificially suppress the impact of energy feedback. Because (at a fixed resolution and heating temperature) the cooling time of a hydrodynamic resolution element decreases more rapidly with increasing density ($\propto n_{\rm H}^{-1}$) than does its sound crossing time ($\propto n_{\rm H}^{-1/3}$), there exists a threshold density above which a resolution element radiates away injected heat before it can expand \citep{Dalla_Vecchia_and_Schaye_12}. At intermediate resolution, approximately half of all stars in the FBconst model form from gas with density above this threshold, resulting in the formation of galaxies that are too old, too passive and too compact. This demonstrates that it is possible to resolve the overcooling problem at the halo scale (i.e. ensuring that haloes form galaxies whose masses are consistent with constraints from, for example, abundance matching techniques) without addressing the angular momentum problem, and hence still yielding a galaxy population that is clearly unrealistic.

The problem is, partially, a consequence of the inability of cosmological simulations to resolve the formation of the first stars. The first galaxies whose formation \textit{can} be captured are necessarily associated with haloes that have not yet been subject to internal feedback processes, exhibiting unrealistically high gas fractions and star formation efficiencies, and potentially triggering a cycle of overestimated radiative losses.  The FB$\sigma$ and FBZ simulations, for which $f_{\rm th}$ increases with decreasing dark matter velocity dispersion and metallicity, respectively, successfully interrupt this cycle by appealing to $f_{\rm th}>1$ for stars forming within nascent galaxies. However, this modification alone is insufficient to inhibit the onset of numerical overcooling. The comoving stellar mass density of these simulations rises much more rapidly than observed at early cosmic epochs, indicating that star formation is initially too efficient, and must decline (too) strongly at later times in order to reproduce the observed stellar mass density, and the GSMF, at $z=0.1$. As in FBconst, galaxies in these simulations exhibit massive, compact bulges that are a classic signature of overcooling. 

The suppression of numerical losses is crucial not only to ensure that the growth of galaxies is appropriately regulated, but also to minimise the risk of misinterpretation. If significant numerical losses are overlooked, the macroscopic efficiency of feedback (as adopted in a particular simulation) will be underestimated. The energy budget then required for outflows to balance inflows, and so to reproduce a realistic galaxy population, will be overestimated. Spurious conclusions are then easily drawn. For example, if numerical losses associated with star formation feedback are overlooked, one might conclude that the shallow faint end of the GSMF can only be reproduced by appealing to non-standard solutions such as a top-heavy IMF, gravitational preheating or the adoption of a warm dark matter (WDM) cosmogony. In the case of more massive galaxies, overlooking numerical losses might lead to the inference that very efficient AGN feedback is necessary to regulate star formation. These conclusions may of course be reasonable, but they should only be inferred from the analysis of simulations with demonstrably insignificant numerical radiative losses. 

The ideal solution to the numerical overcooling problem exhibited by the FBconst, FB$\sigma$ and FBZ simulations is to appeal to higher resolution. The critical density for efficient feedback scales $n_{\rm H,t_c} \propto m_{\rm g}^{-1/2}$. Moreover, the improved sampling afforded by higher resolution justifies the adoption of a higher heating temperature, which implies a lower heating probability per newly-formed star particle. This is an effective means of increasing the critical density, since $n_{{\rm H},t_{\rm c}} \propto T^{3/2}$. The computational cost of higher resolution is currently prohibitive, however, particularly since the calibration of subgrid parameters requires that ``production'' simulations are preceded by a suite of smaller exploratory runs. Experience from the calibration of the EAGLE simulations indicates that volumes of at least $L = 25\cMpc$ are necessary in order to adequately sample the population of galaxies of stellar mass corresponding to the knee of the GSMF, whose properties are particularly sensitive to the subgrid parameters governing feedback from both star formation and AGN. 

Our solution is to adopt a star formation feedback efficiency that increases with density at fixed metallicity. This inhibits the ISM from reaching the high densities at which our stochastic thermal heating scheme becomes numerically inefficient. This critical density, $n_{{\rm H},t_{\rm c}} \sim 10\,\cmcubed$, is much greater than the density at which a cold interstellar phase is expected to develop \citep[$n_{\rm H} \sim 0.1\,\cmcubed$, e.g.][]{Schaye_04}. We do not include such a phase in our simulations, since they lack the resolution required to model it accurately. The simulations treat the mutiphase ISM as a single-phase medium that is subject to a polytropic equation of state; in reality, losses within a cold phase would be significantly higher than those expected from this volume-averaged representation, but much of the energy injected by feedback is likely to be channelled into the hot, diffuse phase of the ISM whose radiative losses will, conversely, be much lower. Moreover, the formation of a cold, clumpy ISM phase would also lead to more clustered star formation, and higher-temperature outflows with relatively low radiative losses. Clearly, the simulations cannot model radiative losses accurately for gas with $n_{\rm H} \gg 0.1\,\cmcubed$, and the adoption of a density dependence for $f_{\rm th}$ can be justified.

This scheme enables the EAGLE reference model to be calibrated to reproduce the GSMF, and the size-mass relation of disc galaxies, as observed at $z=0.1$. This model therefore resolves the overcooling problem at all radii within galaxies, supressing the formation of unrealistically massive bulge components and alleviating the angular momentum problem. S15 further demonstrated that, with the same calibration, many other observed properties of galaxies at $z=0.1$ are reproduced, for example specific star formation rates, passive fractions, BH masses, rotation velocities and metallicities, in addition to the number density of intergalactic absorption systems. \citet{Furlong_et_al_14_short} demonstrated that the EAGLE reference model also reproduces the observed evolution of the comoving stellar mass density and galaxy specific star formation rates.

Integrating over all feedback events associated with star formation, the median subgrid feedback efficiency in the Ref-L050N0752 and Ref-L100N1504 simulations is $f_{\rm th} = 0.65$ and $0.70$, respectively, in spite of this model appealing to $f_{\rm th}>1$ when stars are born from high-density gas. The similarity of these values to the expected energy budget, i.e. that the median efficiency is of order unity, might be considered remarkable. Choosing to \textit{calibrate} the simulations affords the freedom to adopt values much greater or much less than unity if such values are required in order to reproduce the calibration diagnostics. That such values did not prove to be necessary is a non-trivial outcome. It indicates that the radiative losses that shape the properties of the galaxy population and their environments are established primarily on scales that are ``macroscopic'' in the context of cosmological simulations. 

Having calibrated a model that enables the GSMF and sizes of galaxies to be reproduced at $z=0.1$, we have explored the sensitivity of galaxy properties and scaling relations to variation of the key parameters of the subgrid routines. In general, the properties of the galaxy population are insensitive to the adopted star formation threshold. Similarly, when self-regulation is dominated by star formation feedback, the properties of galaxies are largely insensitive to the details of the ISM, as was also found by \citet{Schaye_et_al_10_short} and \citet{Haas_et_al_13b}. The properties of more massive galaxies, whose growth is regulated by AGN feedback, are however sensitive to the assumed equation of state of the ISM, because the accretion rate onto BHs (and hence the liberation of energy as AGN feedback) is inversely proportional to the cube of the local sound speed which, for gas subject to a polytropic equation of state, scales as $c_{\rm s}^2 \propto \rho^{\gamma_{\rm eos}-1}$.

The properties of galaxies are in general sensitive to the value of any parameter that affects energetic feedback processes. In common with the findings of many semi-analytic \citep[e.g.][]{Benson_et_al_03,Bower_et_al_06,Croton_et_al_06_short,Somerville_et_al_08,Bower_Benson_and_Crain_12} and hydrodynamical \citep[e.g.][]{Haas_et_al_13a,Puchwein_and_Springel_13,Vogelsberger_et_al_13,Khandai_et_al_14,Vogelsberger_et_al_14_short} simulations, the properties of galaxies with stellar mass $\lesssim M^\star$ at $z=0.1$ are largely governed by feedback associated with star formation, whilst those of more massive galaxies are governed by feedback associated with the growth of BHs. Feedback regulates the conversion of gas into stars, such that, at fixed halo mass, galaxy stellar masses are lower. More efficient feedback therefore results in the association of galaxies of fixed stellar mass with more massive haloes. It also fosters the formation of more extended galaxies, by preferentially ejecting gas with the low angular momentum from the ISM.

Some consequences of more efficient feedback are less intuitive. For example, more efficient star formation feedback results in higher specific star formation rates at $z=0.1$. This is because galaxies of a fixed stellar mass are hosted by more massive haloes, so experience a greater cosmological accretion rate and must increase their SFR to achieve self-regulation. By extension, more efficient feedback also leads to reduced passive fractions at $z=0.1$. It is also potentially counter-intuitive that more efficient feedback, in spite of associating galaxies of fixed stellar mass with more massive haloes, results in lower central BH masses. This can also be explained in terms of self-regulation: more efficient star formation feedback enables inflows to be balanced without the need for the strong AGN feedback that accompanies rapid BH growth. 

We have not studied simulations in which the subgrid efficiency of AGN feedback, $\epsilon_{\rm f}$, has been varied, since, as demonstrated by \citet{Booth_and_Schaye_09,Booth_and_Schaye_10}, this only affects the mass of BHs. However, effects similar to the variation of $f_{\rm th}$ do accompany the variation of the subgrid viscosity, $C_{\rm visc}$, and the AGN heating temperature, $\Delta T_{\rm AGN}$. A higher viscosity enables BHs to begin accreting efficiently at a lower mass, boosting the mass of BHs at fixed stellar mass, particularly so for the low-mass galaxies whose BHs grow primary by viscosity-limited gas accretion \citep{Rosas_Guevara_et_al_14_short}. It also results in the formation of more extended massive galaxies, since the earlier onset of AGN feedback enables the removal of a greater fraction of low angular momentum gas from their ISM. A higher heating temperature, despite making heating events more intermittent, boosts the efficiency of AGN feedback in a fashion similar to increasing $f_{\rm th}$ for star formation feedback. Because BH mass is established, to first order, by halo mass \citep{Booth_and_Schaye_10}, the resulting shift in the halo mass associated with a galaxies of fixed stellar mass results in a similar shift of BH masses, and fosters the formation of more extended galaxies.

Ultimately, the corroboration of our conclusions requires simulations with sufficient resolution and physical detail to enable ab initio prediction of energy and momentum losses within the multiphase ISM, and the development of such calculations is likely to remain beyond our reach in the near future. Whilst progress is being made in pursuit of this goal \citep[e.g.][]{Yirzak_Frank_and_Cunningham_10,Aluzas_et_al_12}, and valuable attempts are underway to bridge the vast dynamic range between star-forming complexes and galaxies \citep[e.g.][]{Joung_and_Mac_Low_06,Powell_Slyz_and_Devriendt_11,Hopkins_Quataert_and_Murray_12,Agertz_et_al_13,Creasey_Theuns_and_Bower_13,Hopkins_et_al_14,Creasey_Theuns_and_Bower_15}, there are many profitable lines of enquiry to pursue with cosmological simulations that appeal to subgrid models. As was discussed by S15, the chief purpose of such simulations is not to yield ab initio predictions of the masses of galaxies and their BHs, but to illuminate the fashion by which astrophysical processes shape the evolution of galaxies and their environments. Having calibrated simulations to reproduce judiciously chosen observables, their confrontation with myriad complementary observables, many more than explored here, represents an efficient and direct means of identifying shortcomings in galaxy formation theory, and motivating improved treatments of the relevant astrophysical processes.


\section*{Acknowledgements}  
\label{sec:acknowledgements}

We gratefully acknowledge the expert high performance computing support of Lydia Heck and and Peter Draper, and useful discussions with Adam Muzzin and Ivan Baldry. We thank Yannick Bah\'e for a careful reading of the initial manuscript, and the anonymous referee for constructive feedback that improved the article. RAC is a Royal Society University Research Fellow. We thank PRACE for awarding us access to the Curie facility based in France at Tr\`es Grand Centre de Calcul. This study used the DiRAC Data Centric system at Durham University, operated by the Institute for Computational Cosmology on behalf of the STFC DiRAC HPC Facility (www.dirac.ac.uk); this equipment was funded by BIS National E-infrastructure capital grant ST/K00042X/1, STFC capital grant ST/H008519/1, STFC DiRAC Operations grant ST/K003267/1 and Durham University. DiRAC is part of the National E-Infrastructure. The study was sponsored by the Dutch National Computing Facilities Foundation (NCF) for the use of supercomputer facilities, with financial support from the Netherlands Organisation for Scientific Research (NWO), and the European Research Council under the European Union’s Seventh Framework Programme (FP7/2007-2013) / ERC Grant agreements 278594 GasAroundGalaxies, GA 267291 Cosmiway, and 321334 dustygal. Support was also received via the Interuniversity Attraction Poles Programme initiated by the Belgian Science Policy Office ([AP P7/08 CHARM]), the National Science Foundation under Grant No. NSF PHY11-25915, and the UK Science and Technology Facilities Council (grant numbers ST/F001166/1 and ST/I000976/1) via rolling and consolidating grants awarded to the ICC.


\newpage
\bibliographystyle{mn2e}
\bibliography{bibliography} 
\bsp


\label{lastpage}
\end{document}